\documentclass[onecolumn,superscriptaddress,
amsmath,amssymb, aps, pre,]{revtex4-1}
\usepackage{dcolumn}
\usepackage{bm}
\usepackage{graphicx}
\usepackage{comment}
\usepackage[caption=false]{subfig}
\usepackage{color}
\usepackage{enumerate}

\newcommand{\bea}{\begin{eqnarray}}

\newcommand{\eea}{\end{eqnarray}}

\begin{document}

\preprint{}
\title{The number of minima in random landscapes generated by constrained random walk and L\'evy flights: universal properties}
\author{Anupam Kundu}
\address{International Centre for Theoretical Sciences, 
Tata Institute of Fundamental Research, Bengaluru -- 560089, India}
\author{Satya N. Majumdar}
\address{LPTMS, CNRS, Univ. Paris-Sud, Universit\'e Paris-Saclay, 91405 Orsay, France}
\author{Gr\'egory Schehr}
\address{Sorbonne Universit\'e, Laboratoire de Physique 
Th\'eorique et Hautes Energies, CNRS UMR 7589, 4 Place Jussieu, 75252 Paris Cedex 05, France}

\begin{abstract}
We provide a uniform framework to compute the exact distribution of the number of minima/maxima in three different random walk landscape models in one dimension. The landscape is generated by the trajectory of a discrete-time continuous space random walk with arbitrary symmetric and continuous jump distribution at each step. In model I, we consider a ``free'' random walk of $N$ steps. In model II, we consider a ``meander landscape'' where the random walk, starting at the origin, stays non-negative up to $N$ steps. In model III, we study a ``first-passage landscape'' which is generated by the trajectory of a random walk that starts at the origin and stops when it crosses the origin for the first time.   
We demonstrate that while the exact distribution of the number of minima is different in the three models, for each model it is universal for all $N$, in the sense that it does not depend on the jump distribution as long as it is symmetric and continuous. In the last two cases we show that this universality follows from a non trivial mapping to the Sparre Andersen theorem known for the first-passage probability of discrete-time random walks with symmetric and continuous jump distribution. Our analytical results are in excellent agreement with our numerical simulations.
\end{abstract}

\maketitle

\section{Introduction}

A typical realisation of a random landscape in dimension $d \geq 2$ consists of local minima, local maxima and saddle points. How many such stationary 
points are there in a given realisation of the landscape? How do these numbers fluctuate from sample to sample? These are fundamentally important questions
in a wide variety of fields including physics, chemistry, mathematics and computer science~\cite{Adler, Azais, freund1995saddles, halperin1966impurity, broderix2000energy, longuet1960reflection,halperin_82,Ros_Fyo_23}. For example, in liquids or glassy systems, the statistics of such stationary points provide information on the complexity (entropy) of metastable states \cite{bray1980metastable,annibale2003supersymmetric,aspelmeier2004complexity,Satya_Martin,Hivert,Sollich}. The same question also appears in string theory~\cite{aazami2006cosmology,susskind2003anthropic}, random matrix theory and also in the characterisation of the fitness landscapes in evolutionary biology~\cite{barton2005fitness,Krug_review,Krug_fitness1,Krug_fitness2}. Another application appears in optics, where one needs to estimate the number of specular points on a random reflecting surface. Similarly, the intensity of the electric field of speckle laser patterns requires the knowledge of the number of stationary points of a Gaussian random field \cite{longuet1960reflection,halperin_82}. This estimation problem also recently appeared in computer science problems dealing with big data where one wants to perform non-convex optimisation of large dimensional data~\cite{Ganguli_14}. Most of these studies concentrated on calculating just the mean number of stationary points of a random landscape, for instance by using the Kac-Rice formalism~\cite{Rice}. 
However, computing the full distribution of the number of such stationary points has largely remained open, except for uncorrelated landscape. In this latter case, the full distribution can be shown to follow Gaussian statistics via the Central Limit Theorem \cite{Satya_Martin,Hivert,Sollich}. 

Thus computing the full distribution of the number of minima, maxima or saddles in a correlated landscape is a difficult and important problem. In a recent paper \cite{KMS24}, we have computed the full distribution of the number of minima/maxima in a one-dimensional correlated landscape generated by the trail of a discrete-time random walk. More precisely, the random landscape of size $N$ in this model is just the spatial trajectory of a discrete-time random walk of $N$ steps with arbitrary symmetric jump distribution at each step. We found that the distribution of the number of minima $m$, given the size $N$ of the landscape, is completely universal for any $N$, i.e., independent of the jump distribution as long as it is symmetric. This was proved by solving an exact recursion relation. Moreover, in this work, we found, rather strikingly, that the  
distribution of the number of minima is also universal for a different random walk landscape, which we called the ``first-passage landscape''. This first-passage landscape is generated by the trajectory of a discrete-time random walk till its first crossing of the origin. The reasons behind the universality in these two models 
were found to be quite different. In the first-passage landscape model, the universality emerged due to a deep connection \cite{KMS24} between the combinatorial problem of counting the number of local minima of such a landscape and the Sparre Andersen theorem of Markov jump processes known in probability theory \cite{SA1954}.

The purpose of this paper is to provide a unified framework that can be used to derive the universal results for the statistics of the number of minima/maxima in both landscape models, namely the free walk and the first-passage walk. Furthermore, we show that this unified framework can be extended to derive universal results for yet another constrained random walk landscape, namely the ``meander landscape''. This meander landscape is generated by the trajectory of a random walker of $N$ steps that starts at zero and remains nonnegative up to $N$ steps.   
Our unified framework is based on an exact mapping between the statistics of the number of minima of such constrained random walk landscapes and the first-passage property of an effective auxiliary random walk. 
This allows us to derive the full distribution of the number of minima for all $N$ and also proving their universalities for the three classes of random walk landscapes. Our analytical predictions are verified by numerical simulations, finding excellent agreement.

The rest of the paper is organized as follows. In Section \ref{sec:model}, we define the model, the observables of interest and also
provide a short summary of the main results. In Section \ref{sec:mapping}, we establish an exact mapping between the statistics of the 
number of minima and the first-passage property of an auxiliary random walk problem. In Section \ref{sec:free} we use this mapping to compute the 
full distribution of the number of minima for a free random walk of $N$ steps. Section \ref{sec:meander} deals with the exact distribution of a landscape
generated by a random walk meander, using this mapping. Finally, in Section \ref{sec:fp} we show how this mapping can be used to compute exactly the distribution of the number of minima in a first-passage landscape. Section \ref{sec:fp} provides a summary and conclusion.

\section{The model, the observables and the summary of the main results} \label{sec:model}

We consider a discrete-time, continuous-space random walker/L\'evy flight on a line
whose position $x_n$ at step $n$ evolves via the Markov jump process
\begin{equation}
x_n=x_{n-1}+ \eta_n\, , \quad{\rm starting}\,\, {\rm from} \quad x_0=0\, ,
\label{rw_def}
\end{equation}
where $\eta_n$'s are independent and identically distributed (IID) jump lengths, each drawn from 
a symmetric and continuous probability distribution function (PDF) $\phi(\eta)$, which
is normalized to unity, $\int_{-\infty}^{\infty} \phi(\eta)\, d\eta=1$. Because of the symmetry,
the mean jump length vanishes identically, $\langle \eta_n\rangle=0$. If the variance
$\sigma^2= \int_{-\infty}^{\infty} \eta^2\, \phi(\eta)\, d\eta$ is finite, the
walk converges, for large $n$, to the Brownian motion via central limit theorem.
In constrast, for L\'evy flights where $\phi(\eta)\sim |\eta|^{-\mu-1}$ with
the L\'evy index $0<\mu\le 2$, the variance of the jump length is infinite due
to the presence of very long jumps.

In this paper, we consider such a random walk/L\'evy flight of $N$ steps, starting at
the origin, but subjected to different spatial constraints. We study three different constrained
walks: (I) an $N$-step free random walk (II) an $N$-step meander random walk where
the walker is constrained to stay non-negative up to $N$-steps and (III) an $N$-step random walk
that stays positive up to $(N-1)$ steps and crosses the origin to the negative side
exactly at step $N$--we refer to it as 
an $N$-step first-passage walk (see Fig. (\ref{typical_traj.fig})). In each model, we count the number $m$ of local minima in
a trajectory of $N$ steps. Clearly, $m$ is a random variable that fluctuates from 
one trajectory to another in any given model. We are interested in computing the
the distribution of the number of local minima $m$, for fixed $N$, in all the three models 
${\rm I}$, ${\rm II}$ and ${\rm III}$.
More precisely, we define the following three probabilities:
\begin{figure}[t]
\includegraphics[width = 0.32 \linewidth]{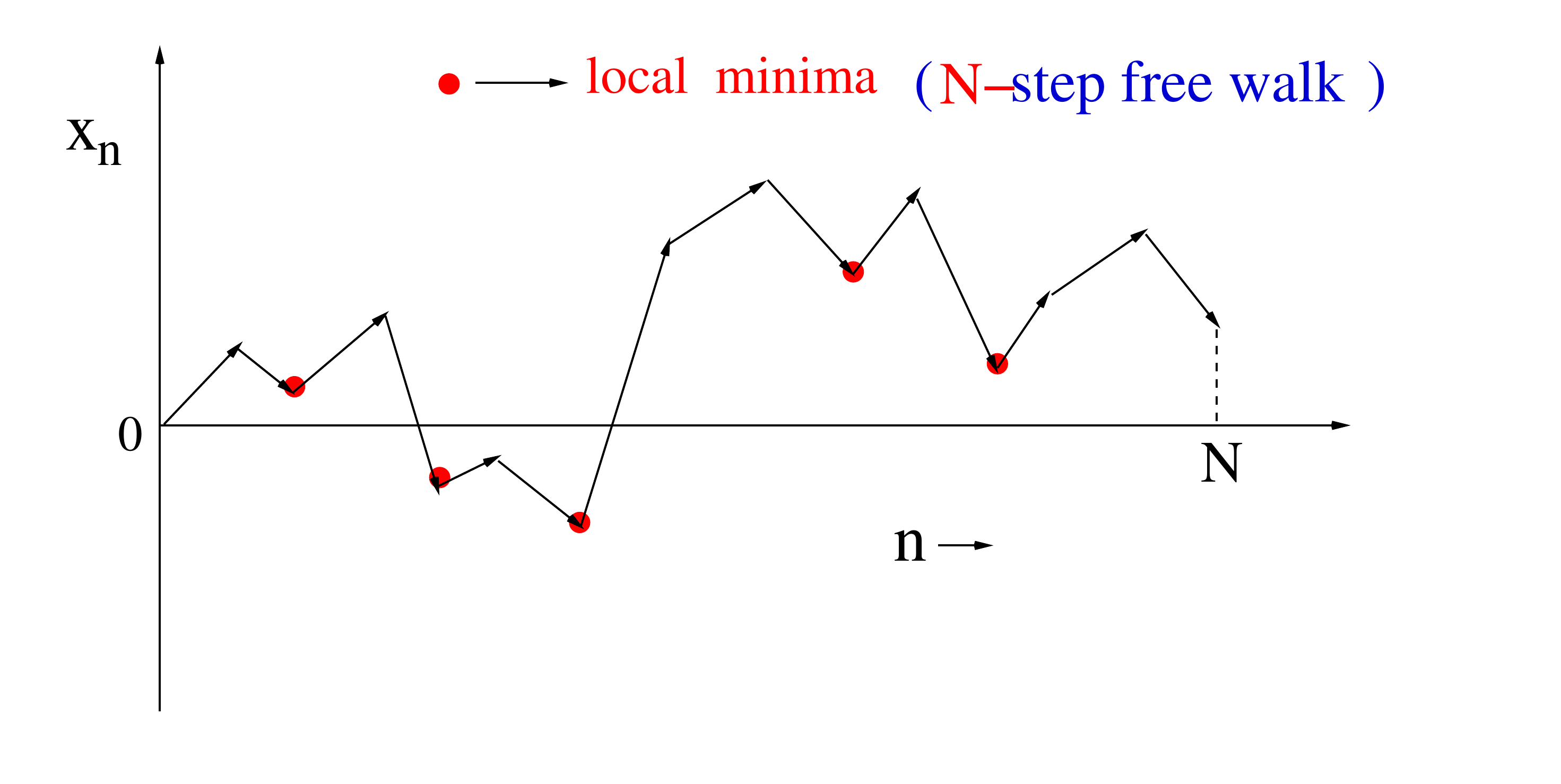}
\includegraphics[width = 0.32 \linewidth]{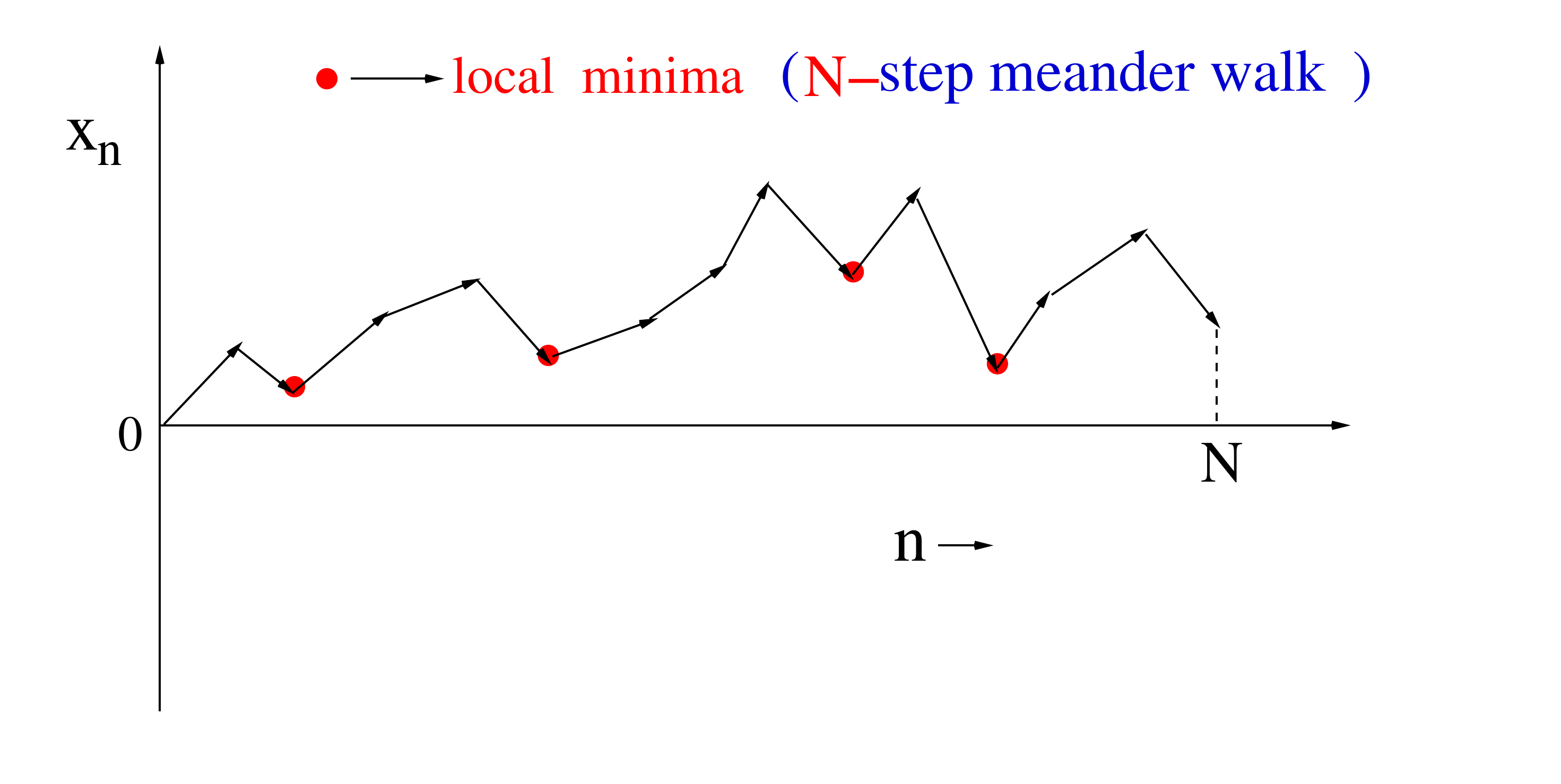}
\includegraphics[width = 0.32 \linewidth]{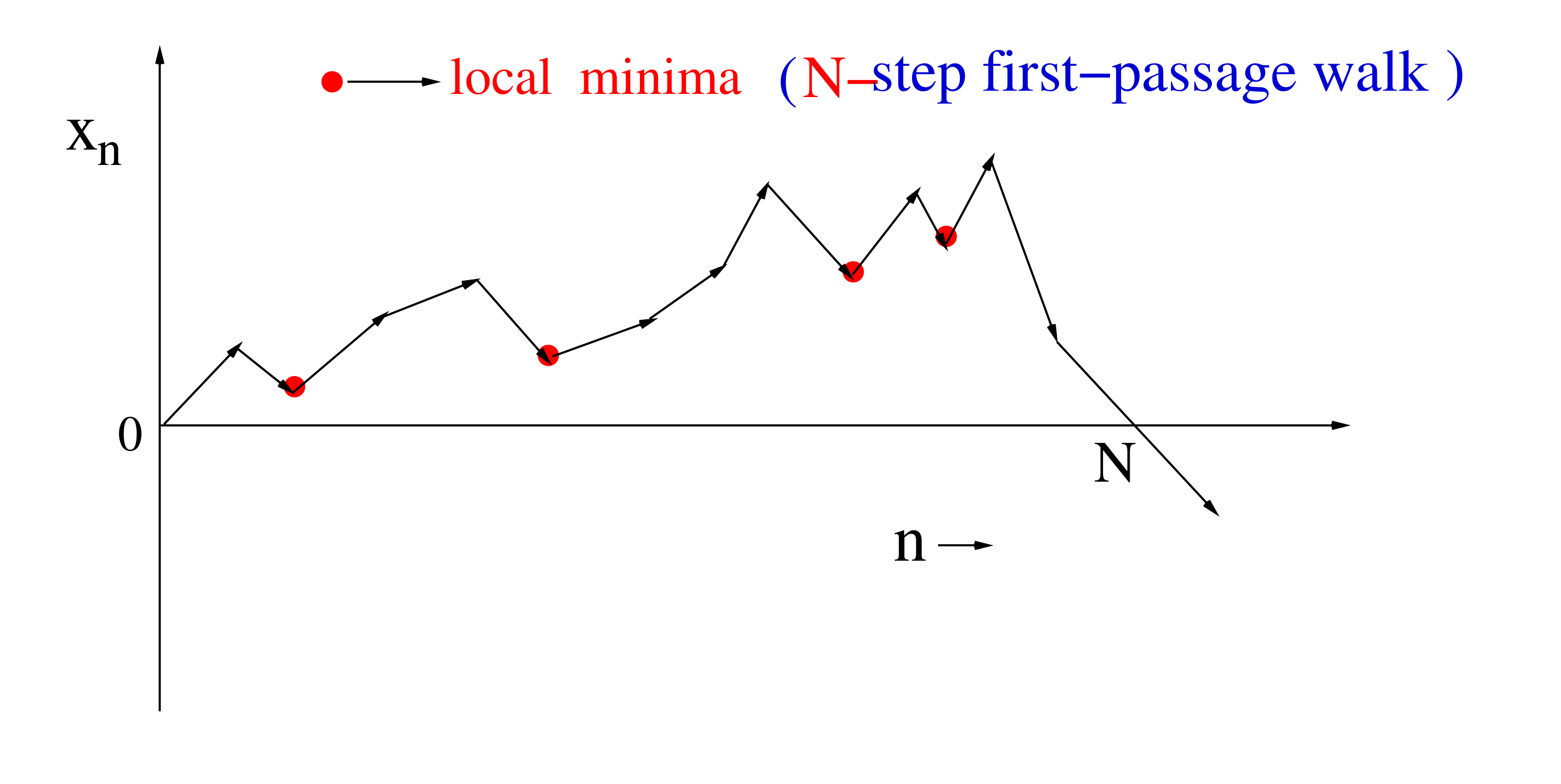}
\caption{(left panel) A typical trajectory of an $N$-step free random walk evolving
via Eq. (\ref{rw_def}) starting from $x_0=0$ ($N=14$). The local minima in this landscape 
are marked by filled
red circles (there are $m=5$ of them). (middle panel) A typical trajectory of an $N$-step meander walk, starting
from $x_0=0$, where
the walker is constrained to stay positive till step $N$ ($N=14$ and $m=4$). 
(right panel) A typical trajectory of a random walk, starting from $x_0=0$ that
stays postive up to step $N-1$ and crosses the origin exactly at step $N$ ($N=14$ and $m=4$). Note that the
origin is not counted as a minimum or maximum for simplicity.}
\label{typical_traj.fig}
\end{figure}

\begin{enumerate}[(I)]

\item $Q(m,N)=$ Probobility of having $m$ local minima in an $N$-step free random walk evolving
via Eq. (\ref{rw_def}), starting from the origin.

\item  $Q_{\rm me}(m,N)=$ Joint probability that the random walk, starting at the origin, 
stays positive up to step $N$ {\em and} that the trajectory has $m$ local minima. The subscript  `me' denotes
a meander.

\item  $Q_{\rm fp}(m, N)=$ Joint probability that the random walk, starting at the origin,
stays positive up to step $(N-1)$ and crosses the origin to the negative side at step $N$
{\em and} that the walk has $m$ local minima. The subscript `fp' stands for a first-passage
walk.

\end{enumerate}

Our main results can be summarized as follows. We show that all three probabilities
listed above are completely universal for all $m$ and all $N$, i.e., independent
of the jump PDF $\phi(\eta)$, as long as it is continuous and symmetric. In Model ({\rm I}),
it does not even need to be continuous. The exact and explicit expressions for
the three probablities are given below.
\begin{itemize}

\item Model {\rm I}. In this case we show that
\begin{eqnarray}
\label{mdist.I}
Q(m,N)  = \begin{cases}
\frac{1}{2^N}\, \frac{(N+1)!}{(N-2m)!\, (2m+1)!} = \frac{1}{2^N} {N+1 \choose 2m+1}\,  \quad {\rm for}\quad 0\le m\le 
\frac{N}{2}   \\
\\
0 \hspace{5.1cm} {\rm for}\quad m>\frac{N}{2} \;.
\end{cases}  
\end{eqnarray}

\item Model {\rm II}. For the $N$-step meander walk, we prove that
\begin{eqnarray}
\label{mdist.II}
Q_{\rm me}(m,N)  = \begin{cases}
 \frac{ (N+1)\, \Gamma(N)}{2^{2m+N+1}\, (m!)^2\, (m+1)\, \Gamma(N-2m)}
\quad\, {\rm for}\quad 0\le m\le \frac{N}{2}-1\, ,    \\
\\
0 \hspace{4.4cm} {\rm for}\quad m> \frac{N}{2}-1, \;. 
\end{cases}
\end{eqnarray}

\item Model {\rm III}. For the $N$-step first-passage walk, we prove that for all $N\ge 2$
\begin{eqnarray}
\label{mdist.III}
Q_{\rm fp}(m,N)  = \begin{cases} \frac{ (N-1)!}{2^{2m+N+1}\, m!\, (m+1)!\, \Gamma(N-2m-1)}
\quad {\rm for}\quad 0\le m\le \frac{N}{2}-1\, ,
   \\
\\
0 \hspace{4.4cm} {\rm for}\quad m > \frac{N}{2}-1\, .
\end{cases}
\end{eqnarray}~For $N=1$, we have $Q_{\rm fp}(m,1)= \frac{1}{2}\, \delta_{m,0}$.

\end{itemize}

We show that all the three exact results in Eqs. (\ref{mdist.I}), (\ref{mdist.II})
and (\ref{mdist.III}) can be derived from the same basic ingredient, namely,  
by exploiting an exact mapping to an auxiliary
random walk that transits from one local minima to the next. We will also see that
while the universality in model I emerges from local properties of the noise variables
(two successive independent jumps in the walk having respectively negative and positive directions
create a local minimum), the mechanism responsible for the universality in models II and III is
much more nontrivial since the actual space (and not just the signatures of local jumps) is
involved. We will see that in these last two models, the universality 
emerges as a consequence of the Sparre Andersen theorem
applied to the auxiliary random walk connecting the local minima.

The result for $Q(m,N)$ for an $N$-step free random walk in Eq. (\ref{mdist.I})
was derived recently in Ref.~\cite{KMS24} by using an alternative method. In addition, the result for
the sum $Q^{\rm fp}(m)= \sum_{N=1}^{\infty} Q_{\rm fp}(m, N)$ was also derived
in Ref.~\cite{KMS24}. However, the results for $Q_{\rm me}(m,N)$ in Eq. (\ref{mdist.II}),
as well as that of $Q_{\rm fp}(m,N)$ in Eq. (\ref{mdist.III}), valid for arbitrary $N$  
(and not just the sum over $N$), are new. Numerical simulations are in excellent agreement
with our analytical predictions in Eqs. (\ref{mdist.I})-(\ref{mdist.III}).

\section{Basic ingredients: an exact mapping to an auxilary random walk}\label{sec:mapping}

\begin{figure}[t]
\includegraphics[width = 0.6\linewidth]{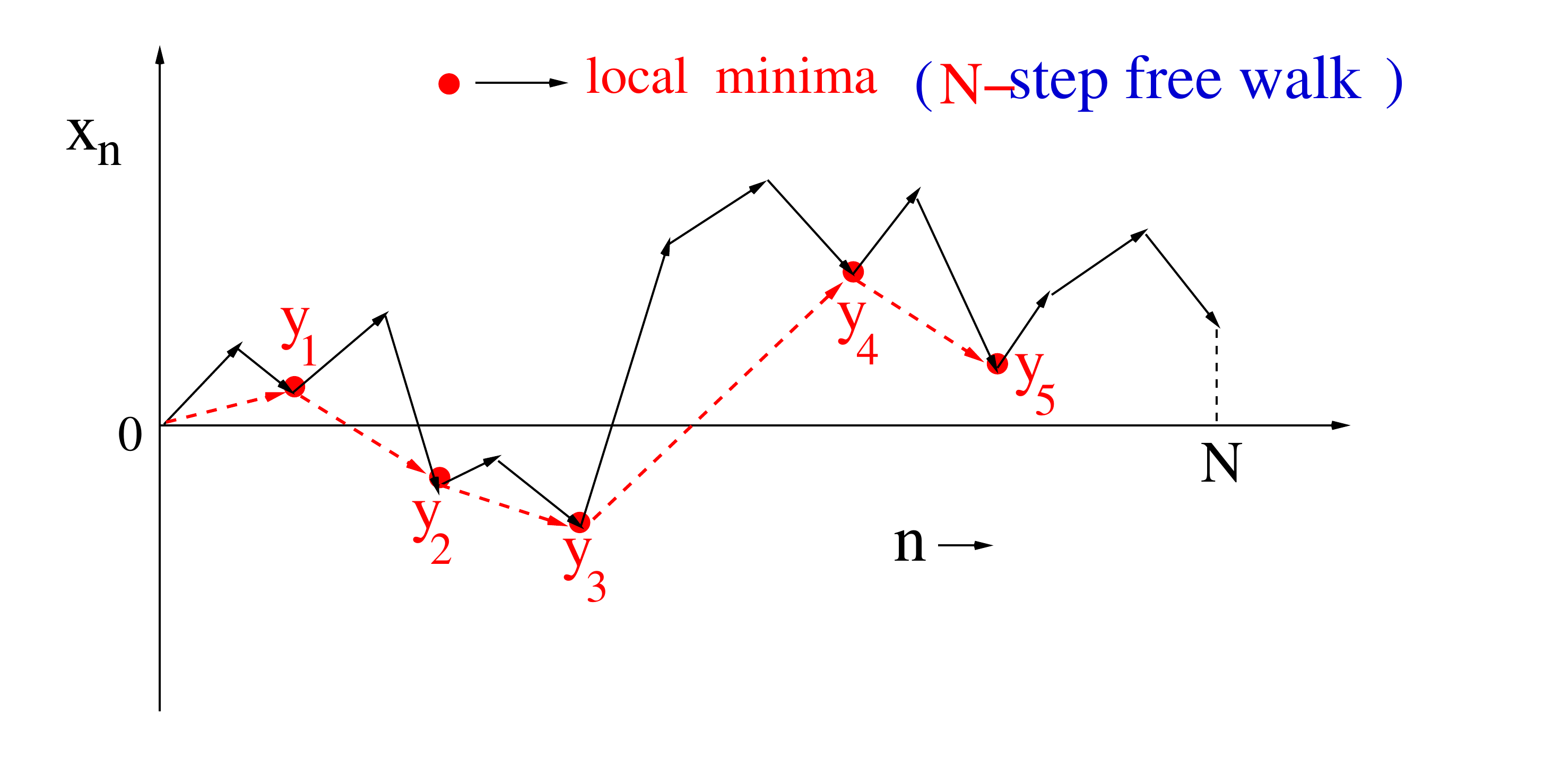}
\caption{A typical trajectory of an $N$-step free random walk evolving
via Eq. (\ref{rw_def}) starting from $x_0=0$ ($N=14$), with the first jump positive. 
The local minima in this landscape
are marked by filled
red circles (there are $m=5$ of them). The dashed red lines connect the origin and all
the local minima and constitute the minima random walk (MRW) trajectory. The red filled circles
(local minima in the original random walk) now denote the positions of the MRW with coordinates
$\{0, y_1,y_2,\ldots, y_m\}$ where $m$ is the number of local minima. The portion of
the trajectory after the last local minima will be referred to as `dangling'.}
\label{MRW1.fig}
\end{figure}

In this section, we will see that the statistics of minima in a random walk landscape generated
via Eq. (\ref{rw_def}), in the presence or the absence of constraints, can be most conveniently
computed by exploiting an exact mapping to an auxiliary random walk problem connecting
only the local minima--we will refer to it as the minima random walk (MRW). To see how this mapping works,
we consider any arbitrary $N$-step trajectory of the random walk generated via Eq. (\ref{rw_def}), as
shown in Fig. (\ref{MRW1.fig}). We locate the local minima and denote their positions by
$\{y_1,y_2,\ldots, y_m\}$ where $m$ is the number of local minima in the
trajectory. We join these local minima by dashed lines and
also connect the origin to the first minima at location $y_1$ (as in Fig. (\ref{MRW1.fig})).
This set $\{0, y_1, y_2, \ldots, y_m\}$ then forms the trajectory
of an auxiliary random walk (we call it MRW), where the MRW jumps from
$y_k$ to $y_{k+1}$ at the $k$-th step. Note that the number of steps of the original
walk between two successive positions of MRW (say between $y_k$ and $y_{k+1}$)
is not fixed and varies from trajectory to trajectory. Our first goal is to
compute the transition probability density $\psi(y_2,y_1,n)$ of the MRW 
to arrive at $y_2$, starting from $y_1$, in $n$ steps of the original random walk.

\begin{figure}\includegraphics[width = 0.6\linewidth]{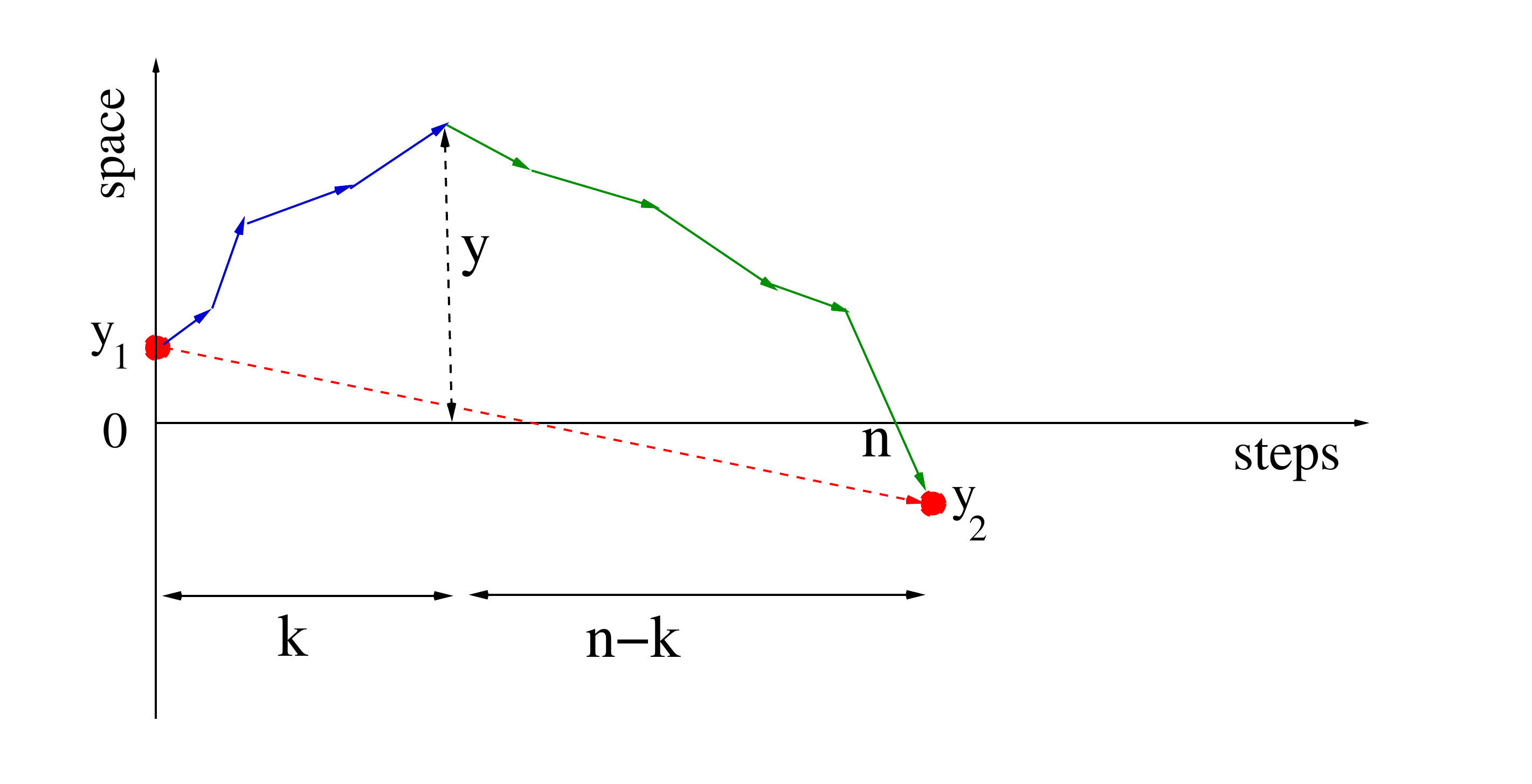}
\caption{The segment consisting of $n$ steps of the orginal walk going from
the local minimum at $y_1$ to the next local minimum at $y_2$.
The segment has a single local maximum with height $y$ that separates
the left part where there are say $k$ jumps that are all positive (marked by blue arrows) and the right part
with $(n-k)$ jumps that are all negative (marked by green arrows).}
\label{MRW_segment.fig}
\end{figure}

To compute the transition probability density $\psi(y_2,y_1,n)$, let us consider the $n$-step
segment of the original walk between two successive local minima with heights $y_1$ and
$y_2$ respectively, as shown in Fig. (\ref{MRW_segment.fig}). The first observation is that
since $y_1$ and $y_2$ are successive local minima by definition, there can not be another
local minumum inside the segment and the number of steps in the segment must be atleast $2$, i.e.,
$n\ge 2$. Also, there must be one and only one local maximum in the interior of this segment
(see Fig. (\ref{MRW_segment.fig})). Clearly, the segment must start from $y_1$ with an upward jump and continue
with upward jumps till it reaches the local maximum, say at height $y$ and then come down to $y_2$
only via negative jumps. The local maximum then divides the segment of $n$ steps into
the left part of upward jumps and a right part of downward jumps. We refer to the left part
as a monotonically upward segment (MUS) and the right part as the monotonically
downward segment (MDS).  Let $k$ and $(n-k)$ denote the number of steps belonging respectively
to the MUS and MDS in a particular configuration (and eventually we will sum over all
allowed values of $k$). Clearly $1\le k\le n-1$. We also note that $y\ge {\rm max}(y_1,y_2)$,
since $y$ is the height of the maximum between $y_1$ and $y_2$. 

Let us first focus on the segment MUS consisting of upward jumps only. Let ${\cal P}_k(x)$ denote
the probability density of reaching $x$, starting from $0$, after $k\ge 1$ successive positive
jumps of the original walk in Eq. (\ref{rw_def}). Then ${\cal P}_k(x)$ satisfies
\begin{equation}
{\cal P}_k(x)= \int_0^x {\cal P}_{k-1}(x')\, \phi(x-x')\, dx'\, , \quad {\rm for}\quad x\ge 0\, ,
\label{Pkx.1}
\end{equation}  
starting from ${\cal P}_0(x)= \delta(x)$. To exploit the convolution structure 
in Eq. (\ref{Pkx.1}),
it is useful to define the Laplace transform
\begin{equation}
\tilde{\cal P}_k(\lambda)= \int_0^{\infty} {\cal P}_k(x)\, e^{-\lambda\, x}\, dx\, .
\label{lap_def.1}
\end{equation}
Consequently, taking Laplace transform of Eq. (\ref{Pkx.1}) and 
using the convolution property, we get
\begin{equation}
\tilde{\cal P}_k(\lambda)= \tilde{\cal P}_{k-1}(\lambda)\, \tilde{\phi}(\lambda)= \left[\tilde{\phi}(\lambda)\right]^k\, ,
\label{Pklam.1}
\end{equation}
where 
\begin{equation}
\tilde{\phi}(\lambda)= \int_0^{\infty} \phi(\eta)\, e^{-\lambda\, \eta}\, d\eta\, .
\label{phi_lam.1}
\end{equation}
Note that since we assume that the jump distribution is symmetric and is normalized to unity
on the whole line, we have from Eq. (\ref{phi_lam.1})
\begin{equation}
\tilde{\phi}(0)= \int_0^{\infty} \phi(\eta)\, d\eta=\frac{1}{2}\, .
\label{phi_norm.1}
\end{equation}
Consequently, from Eq. (\ref{Pklam.1}), we have
\begin{equation}
\int_0^{\infty} {\cal P}_k(x)\, dx= \tilde{\cal P}_k(0)= \left[\tilde{\phi}(0)\right]^k= \frac{1}{2^k}\, .
\label{Pk0.1}
\end{equation}
It is also useful to define the generating function
\begin{equation}
G(x,z)= \sum_{k=1}^{\infty} {\cal P}_k(x)\, z^k \, .
\label{Gxz.1}
\end{equation} 
Integrating over $x\in [0,\infty]$, using (\ref{Pk0.1}) gives
\begin{equation}
\int_0^{\infty} G(x,z)\, dx= \sum_{k=1}^{\infty} \left(\frac{z}{2}\right)^k= \frac{z}{2-z}\, .
\label{Gxz_int.1}
\end{equation} 
Finally, we note that since the jump distribution $\phi(\eta)$ of the original walk is symmetric,
it follows that for the MDS, i.e., the monotonically downward segment on the right of the maximum,
will have the same statistics as the MUS on the left of the maximum. In addition, the two segments
MUS and MDS are statistically independent due to the Markov property of the walk.

Armed with the basic ingredients developed above, we can now write down the transition probability density
$\psi(y_2,y_1,n)$ of the auxiliary random walk MRW connecting successive local minima. Using the independence
of MUS and MDS and the fact that they have the same statistics, we can write
\begin{equation}
\psi(y_2,y_1,n)= \sum_{k=1}^{n-1} \int_{\max(y_1,y_2)}^{\infty} {\cal P}_k(y-y_1)\, {\cal P}_{n-k}(y-y_2)\, dy\, ,
\label{psi_n.1}
\end{equation}
where $y\ge \max(y_1,y_2)$ denotes the height of the maximum between $y_1$ and $y_2$ (see Fig. (\ref{MRW_segment.fig})),
and ${\cal P}_k(x)$ is computed above. Taking generating function with respect to $n$, 
and using (\ref{Gxz.1}), we get
\begin{equation}
\tilde{\psi}_z(y_2,y_1)= \sum_{n=2}^{\infty} \psi(y_2,y_1,n)\, z^n= \int_{\max(y_1,y_2)}^{\infty} G(y-y_1,z)\, G(y-y_2,z)\, dy\, .
\label{psi_z.1}
\end{equation}
Note that the sum over $n$ starts from $n=2$ since the number of steps of the origin walk between two successive
local minima must be at least $2$. Making further a shift $y-{\max({y_1,y_2})}=u$ in the integral in (\ref{psi_z.1}), we get
\begin{equation}
\tilde{\psi}_z(y_2,y_1)= \int_0^{\infty} G(u,z)\, G(u+ |y_1-y_2|, z)\, du\, .
\label{psi_z.2}
\end{equation}
This shows that $\tilde{\psi}_z(y_2,y_1)$ is only a function of the difference $y_2-y_1$, and hence we can write
\begin{equation}
\tilde{\psi}_z(y_1,y_2)= \tilde{\psi}_z(y_2-y_1)\, ,
\label{psi_z.3}
\end{equation}
where $\tilde{\psi}_z(x)$ is a symmetric function of $x$. In addition, this function $\tilde{\psi}_z(x)$ is non-negative for all $x$
and its normalization can be explicitly worked out as follows. Integrating over $x$ and using the symmetry
in (\ref{psi_z.3}) gives
\begin{equation}
\int_{-\infty}^{\infty} \tilde{\psi}_z(x)\, dx = 2\, \int_0^{\infty} du\, G(u,z) \int_0^{\infty} G(u+x,z)\, dx
= 2\, \int_0^{\infty} du\, G(u,z)\,  \int_u^{\infty} G(u', z)\, du'= \left[\int_0^{\infty} G(u,z)\, du\right]^2\, .
\label{psiz_norm.1}
\end{equation}
Finally, using (\ref{Gxz_int.1}) gives
\begin{equation}
\int_{-\infty}^{\infty} \tilde{\psi}_z(x)\, dx= \left(\frac{z}{2-z}\right)^2\, .
\label{psiz_norm.2}
\end{equation}
{Note that this generating function can easily be inverted to obtain
\begin{equation}
\int_{-\infty}^{\infty} \psi(y_1,y_2,n) \, dy_2 = \frac{n-1}{2^n} \;.
\end{equation}
This result can be understood as follows (see Fig. \ref{MRW_segment.fig}): the factor $1/2^n = 1/2^{k} \times 1/2^{n-k}$ corresponds to the probability for the random walk to perform $k$ positive steps followed by $n-k$ negative steps, while the factor $(n-1)$ takes into account the different positions of the maximum, with $1\leq k \leq n-1$.}
It is further useful to define the function
\begin{equation}
w_z(x) = \left(\frac{2-z}{z}\right)^2\, \tilde{\psi}_z(x)\, ,
\label{wz_def}
\end{equation}
such that $w_z(x)$ is a symmetric non-negative function of $x$, parametrized by $z$, and is normalized to unity
\begin{equation}
\int_{-\infty}^{\infty} w_z(x)\, dx= 1\, .
\label{wz_norm.1}
\end{equation}
Thus, to summarize, the generating function of the transition probability density $\psi(y_2,y_1,n)$
can be exressed as
\begin{equation}
\sum_{n=2}^{\infty} \psi(y_2,y_1,n)\, z^n= \tilde{\psi}_z(y_2-y_1)=\left(\frac{z}{2-z}\right)^2\, w_z(y_2-y_1)\, ,
\label{psi_z.4}
\end{equation}
where $w_z(x)$ can be interpreted as a continuous and symmetric PDF normalized to unity.
Eq. (\ref{psi_z.4}) is the main result of this section. We will see in later sections that we can then
use the result in (\ref{psi_z.4}) as the basic building block that will enable us to
compute exactly the statistics of several observables of the MRW. 
Note that $w_z(x)$ depends, of course, on the original jump PDF $\phi(\eta)$, since $\tilde{\psi}_z(y_1,y_2)$ 
depends on $G(x,z)$
as in Eq. (\ref{psi_z.2}) which, in turn, depends explicitly on ${\cal P}_k(x)$ and $\phi(\eta)$ via Eq. (\ref{Gxz.1}).
However, we will see below that we would not need this explicit dependence of $w_z(x)$ on $\phi(\eta)$ to establish
our universal results on the statistics of the number of local minima that do not depend on $\phi(\eta)$. 
Just the fact that $w_z(x)$ can be interpreted as a symmetric, continuous, normalized to unity PDF will be
sufficient for us.

\section{Number of local minima in model I\,: $N$-step free walk} \label{sec:free}

Using the basic building blocks developed in the previous section, here we first apply them to compute
the distribution $Q(m,N)$ of the number of local minima $m$ in an $N$-step free random walk. In order
to use the results of the previous section, it is useful to first consider an extended joint probability
$P(\{y_1,y_2,\ldots, y_m\}, m|N)$ of having $m$ local minima and their positions $\{y_1,y_2,\ldots, y_m\}$
for an $N$-step original free random walk. Once we have this extended joint probability, we can compute the
marginal $Q(m,N)$ by integrating over the $y_i$'s, i.e.,
\begin{equation}
Q(m,N)= \int_{-\infty}^{\infty} dy_1\ldots \int_{-\infty}^{\infty} dy_m\, P(\{y_1,y_2,\ldots, y_m\}, m|N)\, .
\label{free_rw.1}
\end{equation}
To compute the joint probability $P(\{y_1,y_2,\ldots, y_m\}, m|N)$, it is further convenient to break it into
two pieces
\begin{equation}
P(\{y_1,y_2,\ldots, y_m\}, m|N)= P_{+}(\{y_1,y_2,\ldots, y_m\}, m|N)+ P_{-}(\{y_1,y_2,\ldots, y_m\}, m|N)\, ,
\label{Pp_Pm.1}
\end{equation}
where $P_{\pm}(\{y_1,y_2,\ldots, y_m\}, m|N)$ denote the probability of the trajectory that starts with
a positive or negative jump respectively. This amounts to splitting
\begin{equation}
Q(m,n)= Q_+(m,N)+Q_{-}(m,N)\, ,
\label{Qm_split.1}
\end{equation}
where
\begin{equation}
Q_{\pm}(m,N)= \int_{-\infty}^{\infty} dy_1 \ldots \int_{-\infty}^{\infty} dy_m\,  P_{\pm}(\{y_1,y_2,\ldots, y_m\}, m|N)\, .
\label{Qpm.1}
\end{equation}
We will see that these two probabilities $Q_{\pm}(m,N)$ will be diferent.

To proceed, we start by computing the joint probability $P_{+}(\{y_1,y_2,\ldots, y_m\}, m|N)$.
We first consider the event $m\ge 1$. The $m=0$ case is a bit special and will be treated later.
It is easier to compute its generating function 
\begin{equation}
Z_+(\{y_1,y_2,\ldots, y_m\}, m, z)= \sum_{N=2}^{\infty} P_{+}(\{y_1,y_2,\ldots, y_m\}, m|N)\, z^N\, ,
\label{Pp_genf.1}
\end{equation}
that effectively attaches a weight $z$ to each jump of the original random walk. We start with $N=2$ in
(\ref{Pp_genf.1}) since to have a number of local minima $m\ge 1$, the total number of steps $N$ must be
at least $2$.  A typical trajectory contributing to this probability is shown in Fig. (\ref{MRW1.fig}). Using
the mapping to MRW, this trajectory consists of $m$ independent blocks connecting successive local minima and the origin
$\{0\to y_1\to y_2\ldots \to y_m\}$ and then 
a `dangling' segment at the end, after the $m$-th minimum (see Fig. (\ref{MRW1.fig})). This dangling segment can be of two types: (a) either
all jumps in this segement are upwards, i.e., an MUS or (b) this segment may consist of a left segment with
only upward jumps followed by a right segment of only downward jumps. So, we need to compute separately the
weights of this dangling segment, after integrating out the final position at the end of the dangling
segment. In case (a), where only upward jumps occur, the dangling segment contributes
a weight $z/(2-z)$ (this follows from Eq. (\ref{Gxz_int.1})). In case (b), since there is an 
MUS followed by a statistically indepedent
MDS, the corresponding weight factor is just $[z/(2-z)]^2$. Hence, the total weight of the dangling segment to
the generating function is given by
\begin{equation}
{\rm weight}_{\rm dangling}= \frac{z}{2-z} + \left(\frac{z}{2-z}\right)^2= \frac{2z}{(2-z)^2}\, .
\label{dangling_weight.1}
\end{equation}     
Using this result and the independence of $m$ blocks of the MRW, we can then write, for $m\ge 1$,
\begin{eqnarray}
\label{Pp_jpdf.1}
Z_+(\{y_1,y_2,\ldots, y_m\}, m, z)&= & \sum_{N=2}^{\infty} P_{+}(\{y_1,y_2,\ldots, y_m\}, m|N)\, z^N \nonumber \\ 
&=&\frac{2z}{(2-z)^2} \, \frac{z^{2m}}{(2-z)^{2m}}\, w_z(y_1)\, w_z (y_2-y_1)\, w_z(y_3-y_2)\ldots w_z(y_m-y_{m-1})\, ,
\end{eqnarray}
where we used Eq. (\ref{dangling_weight.1}) for the dangling sector and Eq. (\ref{psi_z.4}) for each of the $m$ blocks
of the MRW preceding the dangling sector. We recall that $w_z(x)$ is normalized to unity as in Eq. (\ref{wz_norm.1}).
Now, integrating over $\{y_1,y_2,\ldots, y_m\}$ using (\ref{wz_norm.1}) we get
\begin{equation}
\sum_{N=2}^{\infty} Q_+(m,N)\, z^N =Z_+(m,z)= \int_{-\infty}^{\infty} dy_1\ldots \int_{-\infty}^{\infty} dy_m\,
 Z_+(\{y_1,y_2,\ldots, y_m\}, m, z)= 
\frac{2\, z^{2m+1}}{(2-z)^{2m+2}}\, \quad {\rm for}\quad m\ge 1\, .
\label{Zp_mz.1}
\end{equation}
We now consider the case $m=0$ which is a bit special. In this case, only the dangling segment contributes, and we get, using
Eq. (\ref{dangling_weight.1})
\begin{equation}
\sum_{N=1}^{\infty} Q_+(0,N)\, z^N =  \frac{2z}{(2-z)^2}\, .
\label{Pp_m0.1}
\end{equation}
Now, using this result and the fact that $Q_+(0,1)= 1/2$ (since in one step the probability that 
the walk goes up is exactly $1/2$), we get
\begin{equation}
Z_+(0,z)= \sum_{N=2}^{\infty} Q_+(0,N)\, z^N= \sum_{N=1}^{\infty}Q_+(0,N)\, z^N- \frac{z}{2}= \frac{z^2(4-z)}{2(2-z)^2}\, .
\label{Pp_m0.2}
\end{equation}
Hence summarzing, we get 
\begin{eqnarray}
\label{Zp_mz_result.1}
Z_+(m,z)= \begin{cases}
& \frac{2\, z^{2m+1}}{(2-z)^{2m+2}}\, \quad {\rm for}\quad m\ge 1 \nonumber \\
\\
& \frac{z^2(4-z)}{2(2-z)^2}\,  \quad\quad {\rm for}\quad m=0\, .
\end{cases}
\end{eqnarray}
Let us remark that this result in Eq. (\ref{Zp_mz_result.1}) coincides with Eq. (S18) of Ref.~\cite{KMS24} where
it was derived by a completely different method.   

\begin{figure}[t]
\includegraphics[width = 0.6\linewidth]{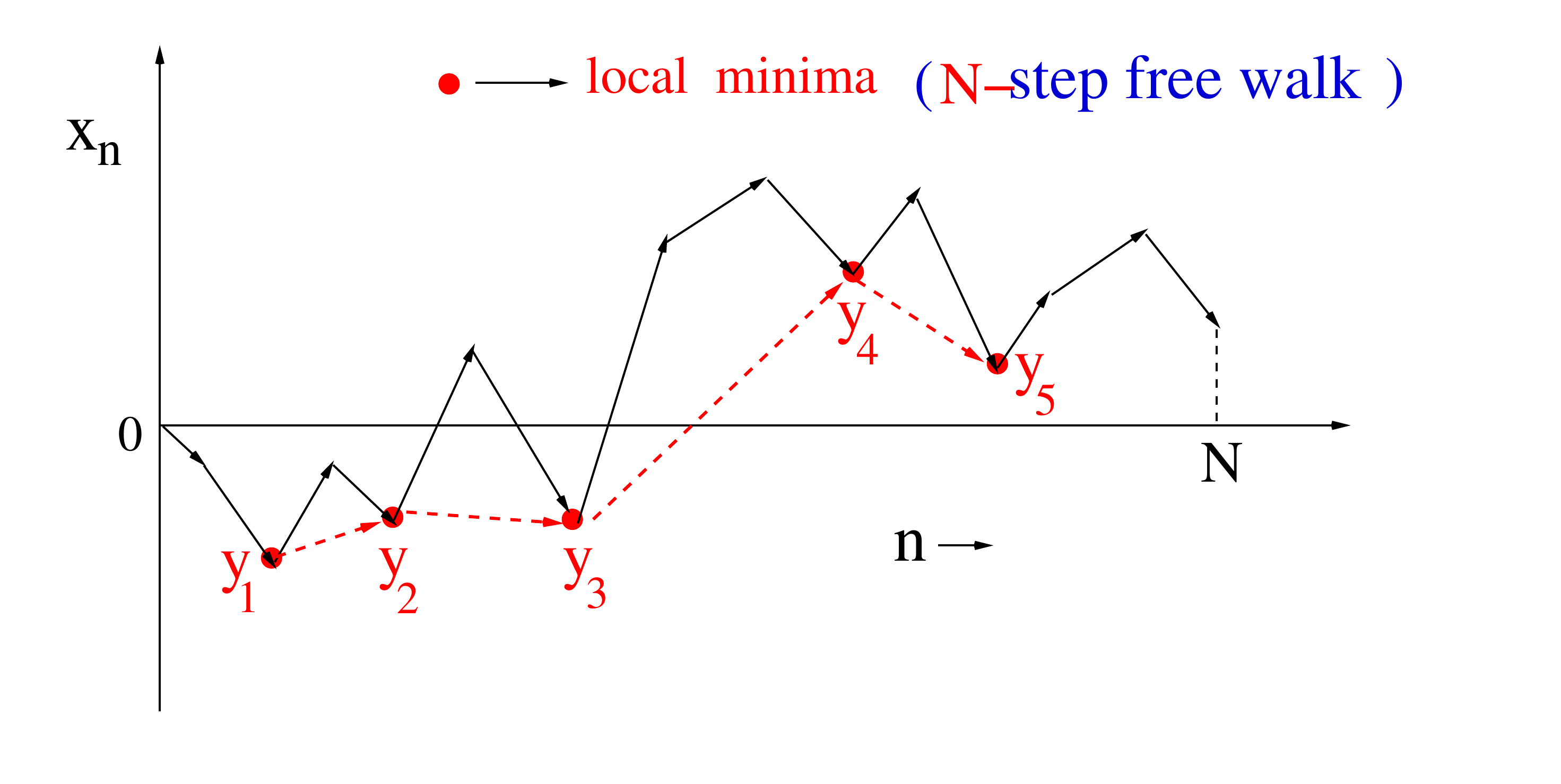}
\caption{A typical trajectory of an $N$-step free random walk evolving
via Eq. (\ref{rw_def}) starting from $x_0=0$ ($N=14$), with the first jump negative.
The local minima in this landscape
are marked by filled
red circles (there are $m=5$ of them), with positions $\{y_1,y_2,\ldots, y_m\}$
where $m$ is the number of local minima. The red dashed lines join
the local minima and constitute the minima random walk (MRW) trajectory starting at $y_1$. 
The red filled circles
(local minima in the original random walk) now denote the positions of the MRW with coordinates
$\{y_1,y_2,\ldots, y_m\}$ where $m$ is the number of local minima. Note that
the origin here is not considered as a part of MRW. The portion of
the trajectory after the last local minima is the `dangling' segment.}
\label{MRW_down.fig}
\end{figure}

We now turn to computing the joint probability $P_{-}(\{y_1,y_2,\ldots, y_m\}, m|N)$, where
the first jump from the origin is negative.
A typical trajectory contributing to  $P_{-}(\{y_1,y_2,\ldots, y_m\}, m|N)$ is shown
in Fig. (\ref{MRW_down.fig}). 
As before, we first consider the event $m\ge 1$ and the $m=0$ case will be treated separately.
As in the previous case, it is easier to compute the generating function
\begin{equation}
Z_{-}(\{y_1,y_2,\ldots, y_m\}, m, z)= \sum_{N=2}^{\infty} P_{-}(\{y_1,y_2,\ldots, y_m\}, m|N)\, z^N\, ,
\label{Pm_genf.1}
\end{equation}
where, by definition, $y_1<0$.
From Fig. (\ref{MRW_down.fig}) it is clear that the trajectory can be broken into $(m-1)$ blocks
connecting the successive local minima $\{y_1,y_2,\ldots, y_m\}$ and the two end segments.
The segment at the end, i.e., after the $m$-th minimum, is just the dangling segment whose
weight was already computed in Eq. (\ref{dangling_weight.1}). The segement at the begining, connecting
the origin to the first local minimum at $y_1$ consists of only downward jumps, starting at the origin
and ending at $y_1$, with $y_1<0$. By using the up-down symmetry of the jump distribution, the weight of this
segment is exactly identical to $G(|y_1|,z)$ where $G(x,z)$ is defined in Eq. (\ref{Gxz.1}). 
Taking the product of the weights of these three parts and using their independence, we can 
then write
\begin{eqnarray}
\label{Pm_jpdf.1}
Z_{-}(\{y_1,y_2,\ldots, y_m\}, m, z)&= & \sum_{N=2}^{\infty} 
P_{-}(\{y_1,y_2,\ldots, y_m\}, m|N)\, z^N \nonumber \\
&=&\frac{2z}{(2-z)^2} \, \frac{z^{2(m-1)}}{(2-z)^{2(m-1)}}\, 
G(|y_1|, z))\, w_z (y_2-y_1)\, w_z(y_3-y_2)\ldots w_z(y_m-y_{m-1})\, ,
\end{eqnarray}
where the dangling sector contributes the factor $2z/(2-z)^2$. We next integrate over $y_1\in (-\infty,0]$
and over $y_k\in (-\infty,\infty)$ for all $1<k\le m$, use Eq. (\ref{Gxz_int.1}) and (\ref{wz_norm.1}), to
get
\begin{equation}
\sum_{N=2}^{\infty} Q_{-}(m,N)\, z^N =Z_{-}(m,z)= \int_{-\infty}^{0} dy_1 \int_{-\infty}^{\infty} dy_2
\ldots \int_{-\infty}^{\infty} dy_m\, Z_-(\{y_1,y_2,\ldots, y_m\}, m, z)=
\frac{2\, z^{2m}}{(2-z)^{2m+1}}\, \quad {\rm for}\quad m\ge 1\, .
\label{Zm_mz.1}
\end{equation}
Let us now consider the case $m=0$, with the first jump negative. In this case, the only configuration that
contributes consists of only downward jumps. Hence we have, trivially, using Eq. (\ref{Gxz_int.1})
\begin{equation}
\sum_{N=1}^{\infty} Q_{-}(0,N)\, z^N= \frac{z}{2-z}\, .
\label{Pm_m0.1}
\end{equation}
Using $Q_{-}(0,1)=1/2$ we then get
\begin{equation}
Z_{-}(0,z)= \sum_{N=2}^{\infty} Q_{-}(0,N)\, z^N= \sum_{N=1}^{\infty} Q_{-}(0,N)\, z^N- 
\frac{z}{2}= \frac{z^2}{2\,(2-z)}\, .
\label{Pm_m0.2}
\end{equation}
Hence summarzing, we get
\begin{eqnarray}
\label{Zm_mz_result.1}
Z_{-}(m,z)= \begin{cases}
& \frac{2\, z^{2m}}{(2-z)^{2m+1}}\, \quad {\rm for}\quad m\ge 1 \nonumber \\
\\
&\frac{z^2}{2\,(2-z)}  \,  \quad\quad\, {\rm for}\quad m=0\, .
\end{cases}
\end{eqnarray}
This result coincides with Eq. (S19) of Ref.~\cite{KMS24} where it was derived by a different method.

\begin{figure}
\centering
\includegraphics[scale=0.2]{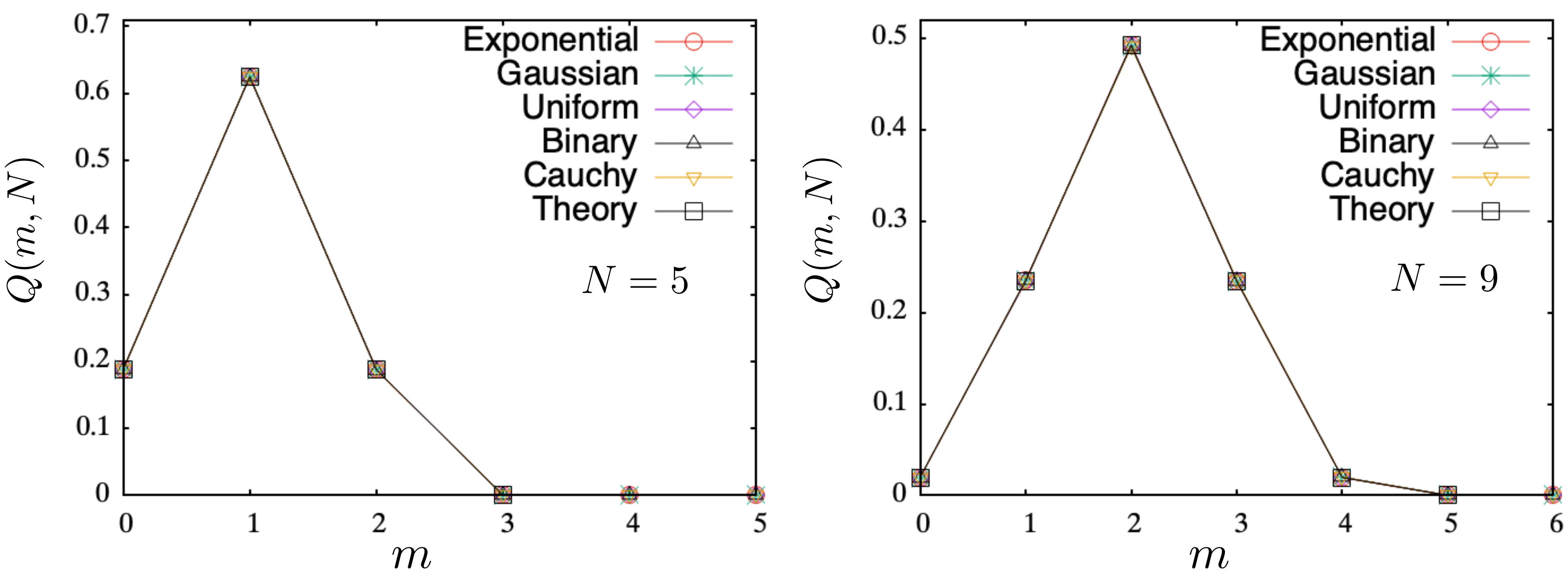}
\caption{Plot of the numerically obtained $Q(m,N)$ for five different jump 
distributions and for two different values of $N$, compared to theoretical expression 
in Eq.~\eqref{result_MI}, plotted with square symbols, showing a perfect agreement. 
The collapse of the data for the different jump distributions on a single curve
clearly demonstrates the universality of $Q(m,N)$.}
\label{QmN_num.fig}
\end{figure}

Taking the generating function of the sum in Eq. (\ref{Qm_split.1})
\begin{equation}
\sum_{N=2}^{\infty} Q(m,N)\, z^N= Z(m,z)= Z_+(m,z)+Z_{-}(m,z) \, ,
\label{Qm_genf.1}
\end{equation}
and adding Eqs. (\ref{Zp_mz_result.1}) and (\ref{Zm_mz_result.1}) we get
\begin{eqnarray}
\label{Z_mz_result.1}
\sum_{N=2}^{\infty} Q(m,N)\, z^N= Z(m,z)= \begin{cases}
& \frac{4\, z^{2m}}{(2-z)^{2m+2}}\, \quad {\rm for}\quad m\ge 1 \nonumber \\
\\
&\frac{z^2 (3-z)}{(2-z)^2}  \,  \quad\quad\, {\rm for}\quad m=0\, .
\end{cases}
\end{eqnarray}
Expanding the right hand side (rhs) of Eq. (\ref{Z_mz_result.1}) in powers of $z$ using the
identity 
\begin{equation}
\left(1-\frac{z}{2}\right)^{-\alpha}= \frac{1}{\Gamma(\alpha)}\, 
\sum_{k=0}^{\infty} \frac{\Gamma(\alpha+k)}{2^k \Gamma(k+1)}\, z^k \, ,
\label{iden.2}
\end{equation}
and matching powers of $z^N$ on both sides gives
\begin{equation}
Q(m,N)= \frac{\Gamma(N+2)}{2^N\, \Gamma(2m+2)\, \Gamma(N-2m+1)}\, \quad {\rm for}\quad 0\le m\le \frac{N}{2}\, ,
\label{result_MI}
\end{equation}
thus completing the derivation of the result announced in Eq. (\ref{mdist.I}). 
Clearly $Q(m,N)$ is completely universal, i.e., independent of the jump distribution $\phi(\eta)$,
 as long as $\phi(\eta)$ is symmetric. In Fig. (\ref{QmN_num.fig}) we compare our theoretical
prediction (\ref{result_MI}) with direct numerical simulations, finding excellent agreement. 
As mentioned earlier,
this result in Eq. (\ref{result_MI}) was already derived in Ref.~\cite{KMS24} by a different method. 
But here we provide a derivation
based on a more general method that can be easily extended to compute the distribution
of the number of local minima for other constrained random walks, as shown in the next two sections. {We also note that this result in
(\ref{result_MI}) was derived in \cite{Marckert} for lattice random walks (i.e., with $\pm 1$ jumps), but here we have shown that it is more general and
holds for random walks with arbitrary symmetric jump distribution}.

\section{Number of local minima in model II\, : $N$-step meander walk} \label{sec:meander}

\begin{figure}[t]
\includegraphics[width = 0.45 \linewidth]{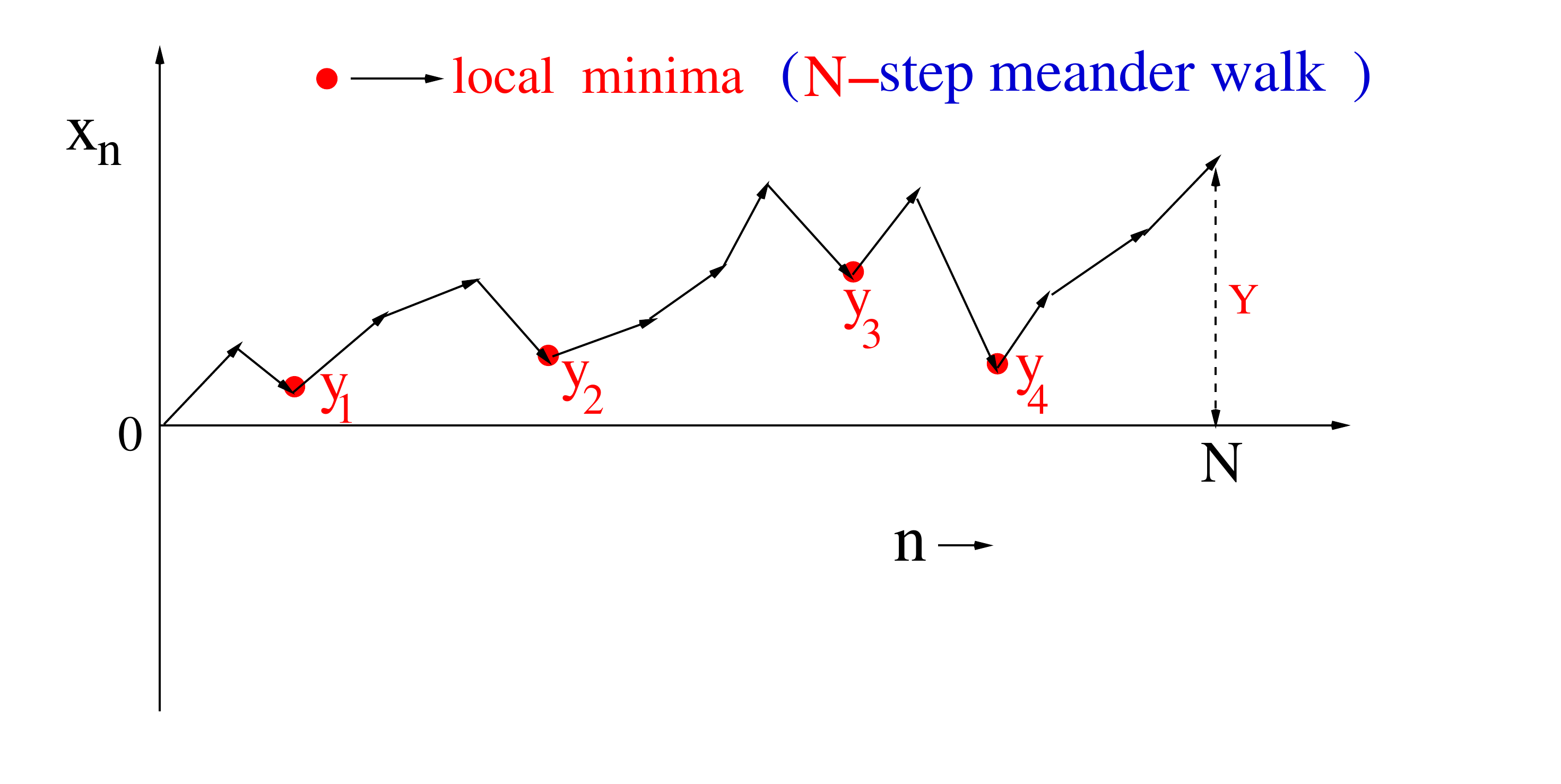}
\includegraphics[width = 0.45 \linewidth]{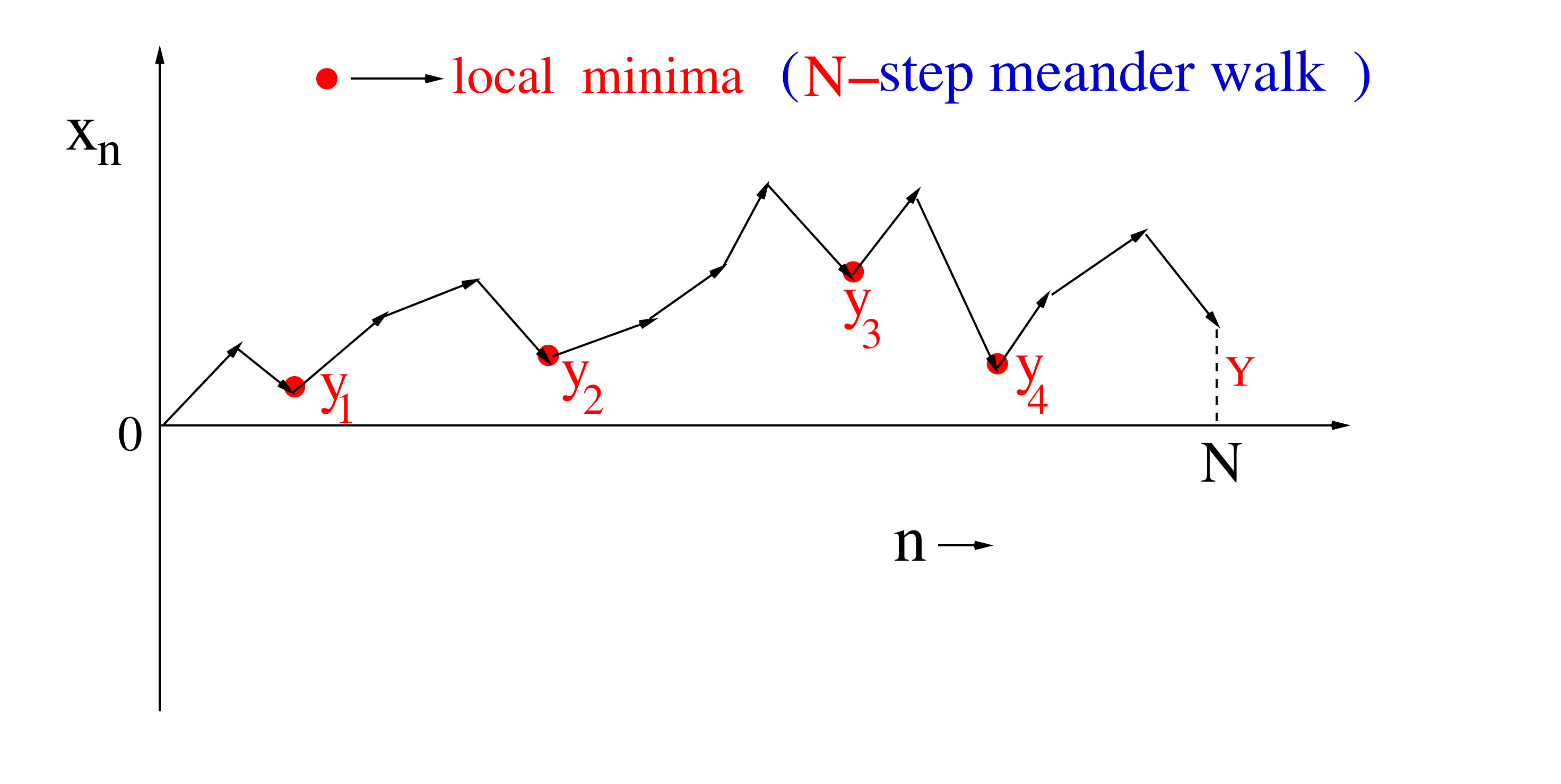}
\caption{Two typical trajectories of an $N$-step ($N=14$) meander random walk
where the walk, starting from the origin, stays non-negative up to step $N$.
(left panel) A meander trajectory where the  last jump is positive and 
(right panel) a meander trajectory where the last jump is negative.
The local minima in this landscape
are marked by filled red circles (there are $m=4$ of them) with coordinates
$\{y_1,y_2,\ldots, y_m\}$. The portion of the walk after the $m$-th minimum
is `dangling' with the coordiante of its end point denoted by $Y$.}
\label{meander_updown.fig}
\end{figure}

In this section we compute $Q_{\rm me}(m,N)$ denoting the probability that the random walk in Eq. (\ref{rw_def})
remains non-negative up to step $N$ and that it has $m$ local minima. 
As in the previous section, it is convenient
to first compute the extended joint distribution of the positions of the local minima and their number and then
integrate out the positions of the local minima. But for the meander, one needs to be careful about the
`dangling' segment, i.e., the portion of the walk beyond the $m$-th minimum. The last jump of this dangling sector can be
either positive (as in the left panel of Fig. (\ref{meander_updown.fig})) or 
negative (as in the right panel
of Fig. (\ref{meander_updown.fig})). One also needs to keep track of the position $Y$ of the last point of the walk 
(see Fig. (\ref{meander_updown.fig})). We first split the probability of having $m$ minima into
two parts depending on whether the last jump is positive or negative
\begin{equation}
Q_{\rm me}(m,N) = Q_{\rm me}^+(m,N) + Q_{\rm me}^{-}(m,N)\, .
\label{Qme_split.1}
\end{equation}
Similarly, we split the  extended joint probability accordingly
\begin{equation}
P(\{y_1,y_2,\ldots y_m\}, Y, m|N)= P^{+}(\{y_1,y_2,\ldots y_m\},Y,m|N)+  P^{-}(\{y_1,y_2,\ldots y_m\},Y,m|N)\, ,
\label{Pjoint_me.1}
\end{equation}
such that
\begin{eqnarray}
Q_{\rm me}^+(m,N) &= & \int_0^{\infty}dy_1\ldots \int_0^{\infty} dy_m\, \int_{y_m}^{\infty} dY\, P^{+}(\{y_1,y_2,\ldots y_m\},Y,m|N)
\label{Qmep.1} \\
Q_{\rm me}^{-}(m,N) &= & \int_0^{\infty}dy_1\ldots \int_0^{\infty} dy_m\, 
\int_0^{\infty} dY\, P^{-}(\{y_1,y_2,\ldots y_m\},Y, m|N)\, .
\label{Qmem.1}
\end{eqnarray}
We remark that the variables $\{y_1,y_2,\ldots, y_m\}$ are integrated in the semi-infinite domain $[0,\infty]$
since we want the walk to remain non-negative.
We note that in Eq. (\ref{Qmep.1}), the limit of integration over $Y$ is over $[y_m,\infty]$. This s because, the
dangling segment in this case is monotonically increasing and hence we must have $Y\ge y_m$. 
In contrast,
in Eq. (\ref{Qmem.1}), $Y$ can take any value in $[0,\infty]$, since the dangling segment 
now consists of an
MUS followed by an MDS.

We start by writing down the expression for $P^{+}(\{y_1,y_2,\ldots y_m\},Y,m|N)$, or rather for its generating function
\begin{equation}
\tilde{P}^+(\{y_1,y_2,\ldots, y_m\}, Y, z)=  \sum_{N=1}^{\infty}
P^{+}(\{y_1,y_2,\ldots, y_m\}, Y,m|N)\, z^N\, .
\label{Pp_genf_me.1}
\end{equation}
From the left panel of
Fig. (\ref{meander_updown.fig}), it is clear that the weight of the dangling segment is an MUS and hence
is given by $G(Y-y_m,z)$, where $G(x,z)$ is defined in Eq. (\ref{Gxz.1}). Then, we have
\begin{equation}
\tilde{P}^+(\{y_1,y_2,\ldots, y_m\}, Y, z)= \left(\frac{z}{(2-z)}\right)^{2m}\, w_z(y_1)\, 
w_z (y_2-y_1)\, w_z(y_3-y_2)\ldots w_z(y_m-y_{m-1})\, G(Y-y_m,z)\, ,
\label{Pp_genf_me.2}
\end{equation}
Consequently, by taking the generating function of Eq. (\ref{Qmep.1}), and using the result $\int_{y_m}^{\infty} 
G(Y-y_m,z)\, dY= z/(2-z)$ from Eq. (\ref{Gxz_int.1}), we get
\begin{equation}
\sum_{N=1}^{\infty} Q_{\rm me}^+(m,N)\, z^N=  \left(\frac{z}{2-z}\right)^{2m+1}\, q_m\, ,
\label{Pp_genf_me.3}
\end{equation}
where we have defined
\begin{equation}
q_m = \int_0^{\infty} dy_1\, \int_0^{\infty} dy_2\, \ldots \int_0^{\infty}d{y_m}\,  w_z(y_1)\, 
w_z (y_2-y_1)\, w_z(y_3-y_2)\ldots w_z(y_m-y_{m-1})\, .
\label{qm_def.1}
\end{equation}
Note, however, that since $w_z(x)$ can be interpreted as the symmetric, continuous and normalized PDF of the
jumps of the MRW (parametrized by $z$), the quantity $q_m$ is just the survival probablity of the MRW, starting at the origin, up to
$m$ steps. By the celebrated Sparre Andersen theorem~\cite{SA1954}, this survival probability $q_m$ is completely
universal, i.e., independent of $w_z(x)$ and is given by the formula
\begin{equation}
q_m= {2m\choose m}\, 2^{-2m} \quad {\rm for}\quad m=0,1,2,\ldots\, .
\label{SA_result.1}
\end{equation}

\begin{figure}
\centering
\includegraphics[width=0.6\linewidth]{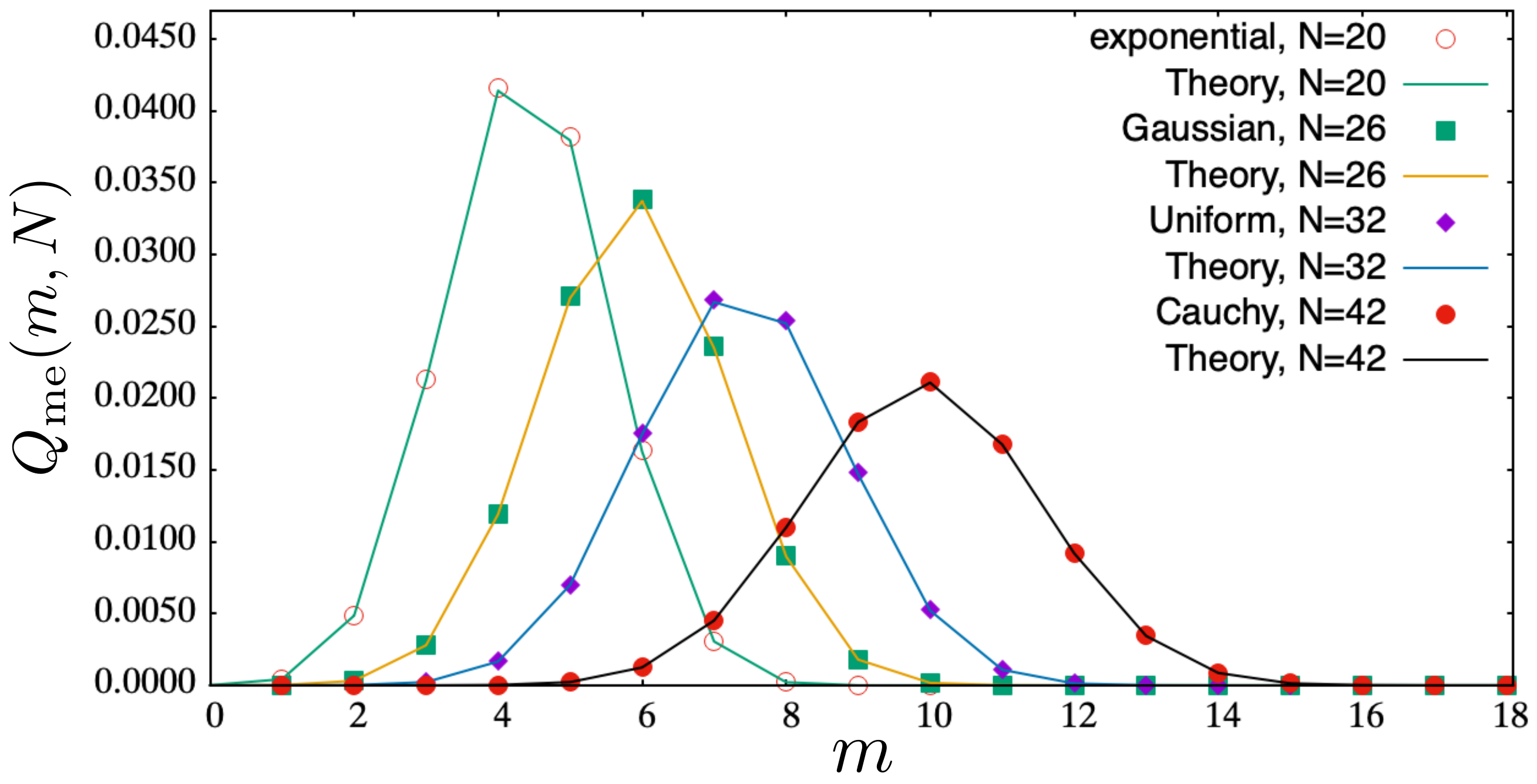}
\caption{Plot of the numerically obtained $Q_{\rm me}(m,N)$ for four different jump
distributions and for four different values of $N$, compared to theoretical expression
in Eq.~\eqref{result_MII} (shown by solid lines), showing a perfect agreement.
}
\label{QmeN_num.fig}
\end{figure}

We now turn to the right panel of Fig. (\ref{meander_updown.fig}) where the last jump is negative.
We define the generating function
\begin{equation}
\tilde{P}^{-}(\{y_1,y_2,\ldots, y_m\}, Y, z)=  \sum_{N=1}^{\infty}
P^{-}(\{y_1,y_2,\ldots, y_m\}, Y,m|N)\, z^N\, .
\label{Pm_genf_me.1}
\end{equation}
From the right panel of
Fig. (\ref{meander_updown.fig}), it is clear that the dangling segment now consists of a monotonically
increasing stretch followed by a monotonically decreasing stretch ending at $Y\ge 0$. Thus the
dangling sector is very similar to any other interior block connecting $y_m$ and $Y$ and 
hence carries a weight $z^2/(2-z)^2\, w_z(Y-y_m)$. Consequently, we get
\begin{equation}
\tilde{P}^{-}(\{y_1,y_2,\ldots, y_m\}, Y, z)= \left(\frac{z}{2-z}\right)^{2\,(m+1)}\, w_z(y_1)\,
w_z (y_2-y_1)\, w_z(y_3-y_2)\ldots w_z(y_m-y_{m-1})\, w_z(Y-y_m, z)\, ,
\label{Pm_genf_me.2}
\end{equation}
Taking the generating function of Eq. (\ref{Qmem.1}) and integrating over the variables in the 
semi-infinite domain $[0,\infty]$ we get
\begin{equation}
\sum_{N=1}^{\infty} Q_{\rm me}^{-}(m,N)\, z^N=  \left(\frac{z}{2-z}\right)^{2m+2}\, q_{m+1}\, ,
\label{Pm_genf_me.3}
\end{equation} 
where $q_m$ is defined in Eqs. (\ref{qm_def.1}) and (\ref{SA_result.1}).

Adding Eqs. (\ref{Pp_genf_me.3}) and (\ref{Pm_genf_me.3}) gives
\begin{equation}
\sum_{N=1}^{\infty} Q_{\rm me}(m,N)\, z^N= \left(\frac{z}{2-z}\right)^{2m+1}\, q_m+ 
\left(\frac{z}{2-z}\right)^{2m+2}\, q_{m+1}\, \quad {\rm for} \, {\rm all}\quad m\ge 0 \, ,
\label{Qme_genf.1}
\end{equation}
where $q_m$ given in (\ref{SA_result.1}) is universal for all $m\ge 0$.
This proves that $Q_{\rm me}(m,N)$ is also universal, i.e., does not depend on the
original jump distribution $\phi(\eta)$, as long as it is symmetric and continuous.
Expanding the rhs of Eq. (\ref{Qme_genf.1}) in powers of $z$ using the identity
(\ref{iden.2}) and reading off the coefficient of $z^N$, it is easy to see that
for $N\ge 2m+1$
\begin{equation}
Q_{\rm me}(m,N)= \frac{q_m\, \Gamma(N)}{2^N\, \Gamma(2m+1)\, \Gamma(N-2m)}+ 
\frac{q_{m+1}\, \Gamma(N)}{2^N\, \Gamma(2m+2)\, \Gamma(N-2m-1)}\, ,
\label{Qme_res.1}
\end{equation}
while $Q_{\rm me}(m,N)=0$ for $N< 2m+1$. Using $q_m$ from Eq. (\ref{SA_result.1}) and simplifying,
we then get our final result announce in Eq. (\ref{mdist.II}), namely
\begin{eqnarray}
\label{result_MII}
Q_{\rm me}(m,N)  = \begin{cases}
 \frac{ (N+1)\, \Gamma(N)}{2^{2m+N+1}\, (m!)^2\, (m+1)\, \Gamma(N-2m)}
\quad\, {\rm for}\quad 0\le m\le \frac{N}{2}-1\, ,    \\
\\
0 \hspace{4.4cm} {\rm for}\quad m> \frac{N}{2}-1, .
\end{cases}
\end{eqnarray}
This analytical result is compared to numerical simulation in Fig. (\ref{QmeN_num.fig}), finding
excellent agreement.

\vspace*{0.5cm}


\section{Number of local minima in model III: first-passage walk} \label{sec:fp}

In this section we compute $Q_{\rm fp}(m,N)$ denoting the joint distribution
that the random walk in Eq. (\ref{rw_def}) crosses the origin from the positive
side for the first time at step $N$ and that it has $m$ minima till this first 
crossing. For a typical trajectory contributing to $Q_{\rm fp}(m,N)$, see the
right panel in Fig. (\ref{typical_traj.fig}). We will see below that this probability
can be related to the probabilities $Q_{\rm me}^{\pm}(m,N)$ for an $N$-step
meander walk defined in Eqs. (\ref{Qmep.1}) and (\ref{Qmem.1}) and 
whose generating functions are explicitly computed respectively in Eqs. (\ref{Pp_genf_me.3})
and (\ref{Pm_genf_me.3}).

\begin{figure}[t]
\includegraphics[width = 0.45 \linewidth]{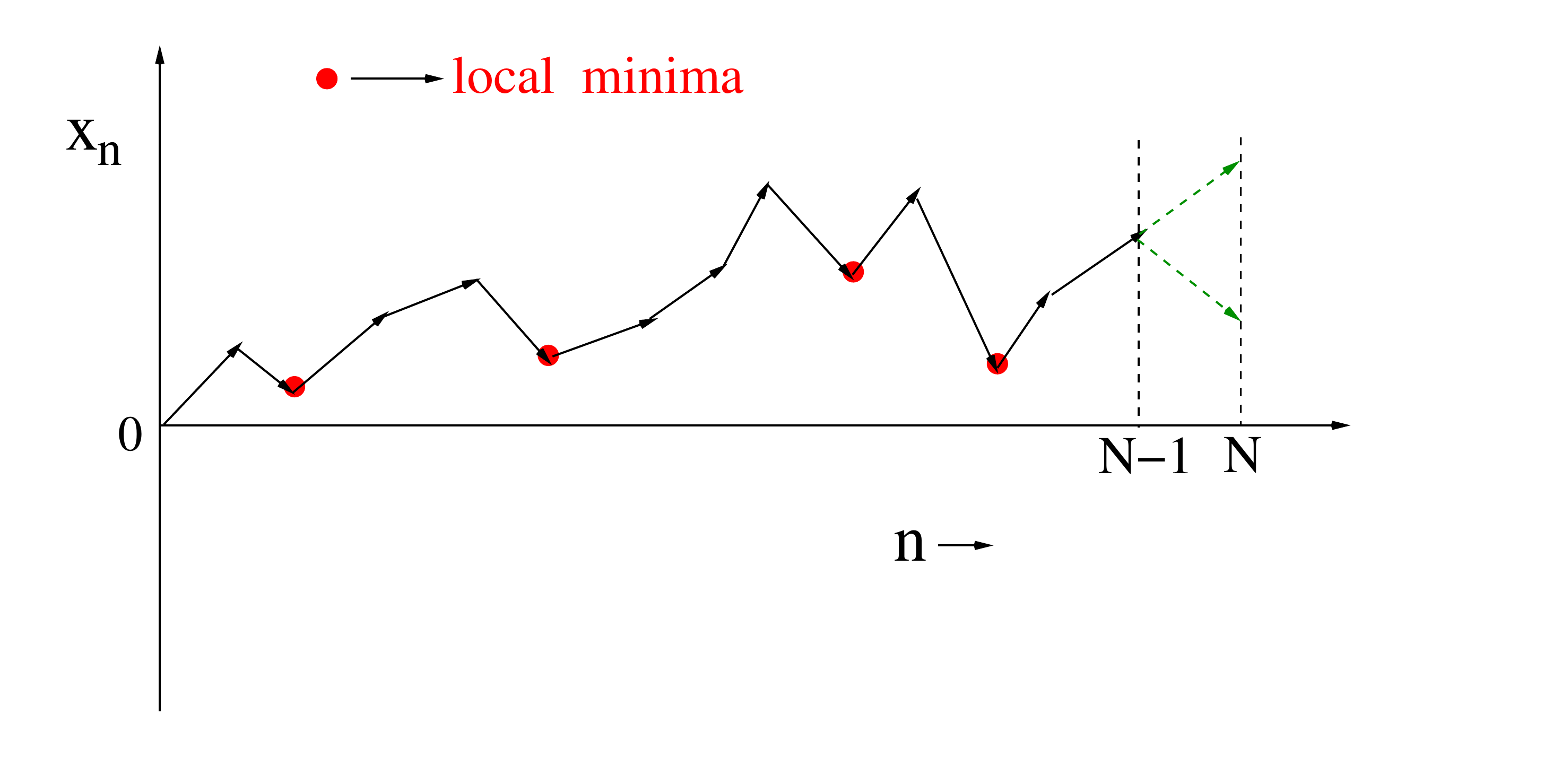}
\includegraphics[width = 0.45 \linewidth]{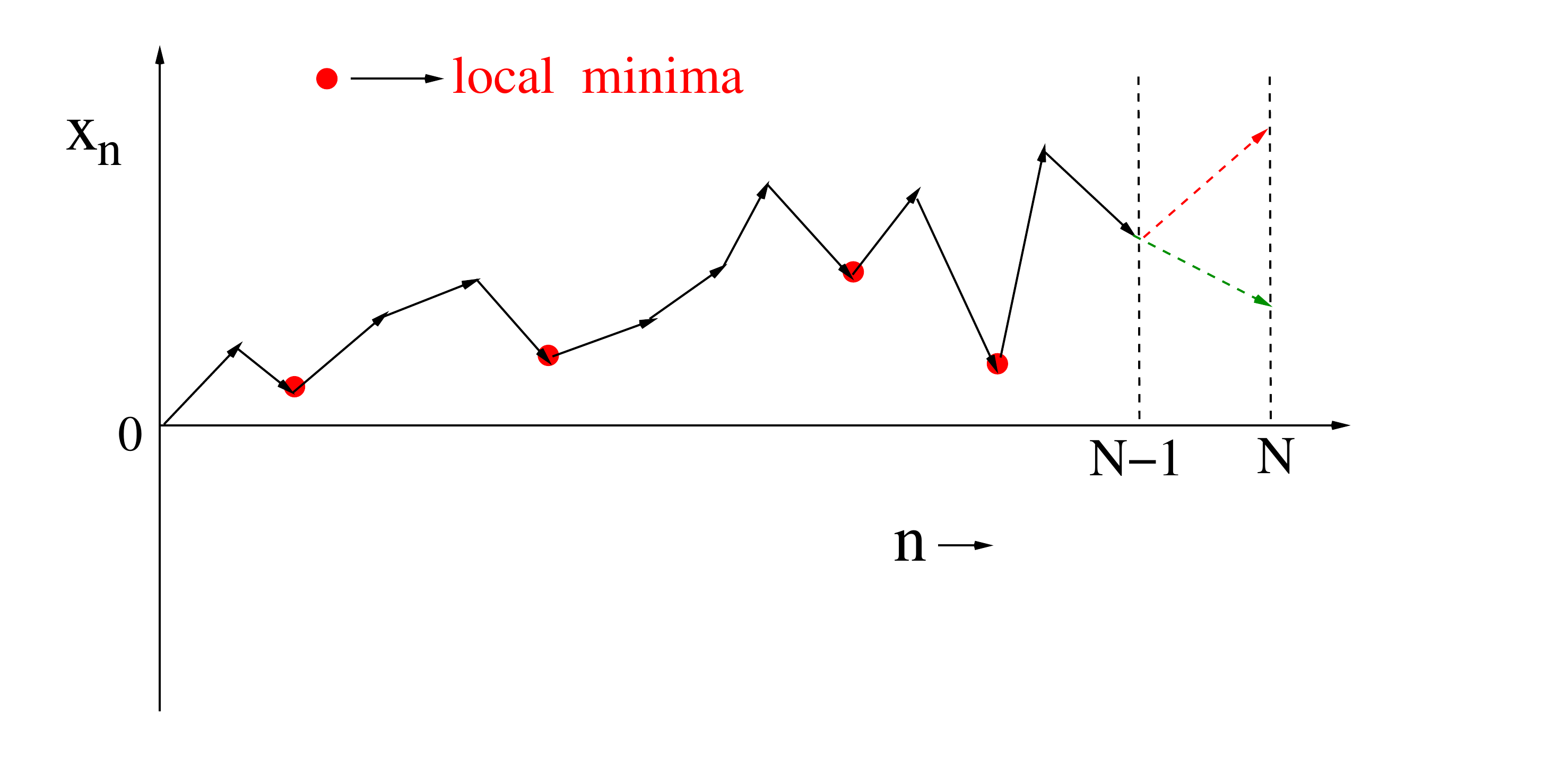}
\caption{(left panel) A trajectory of an $N$-step walk that stays positive up to $(N-1)$ steps
with the $(N-1)$-th jump positive. The $N$-th jump (shown by green dashed arrows) may be
either positive or negative, but in either case it does not generate a new minimum
at the end of the $(N-1)$-th step. (right panel) A trajectory of an $N$-step walk that 
stays positive up to $(N-1)$ steps with the $(N-1)$-th jump negative. The $N$-th jump,
if negative (shown by a green dashed arrow) does not generate a minimum at the end of the
$(N-1)$-th step. However, if the $N$-th jump is positive (as shown by a dashed red arrow),
it does generate a new minimum at the $(N-1)$-th step.}
\label{fp_updown.fig}
\end{figure}

In order to make this connection to meander random walk discussed in the previous section
we proceed as follows. Consider a typical trajectory contributing to $Q_{\rm fp}(m,N)$
in the right panel of Fig. (\ref{typical_traj.fig}). This walk clearly has to stay positive
up to step $(N-1)$ and then has to cross to the negative side at step $N$. We recall
that $Q_{\rm me}^{\pm}(m,N-1)$ denotes the probability that the walk 
stays positive up to step $(N-1)$ (with the last jump positive or negative) and that
it has $m$ minima. Also, $Q_{\rm me}(m,N)= Q_{\rm me}^{+}(m,N)+ Q_{\rm me}^{-}(m,N)$.
Now, it is instructive to see what happens at the $N$-th step. There are two possibilities.
A walk that stays
positive up to the $(N-1)$-th. step, may either stay positive after the $N$-th jump,
or may cross the origin exactly at the $N$-th step. 
Adding these two possibilities, 
we note that
$Q_{\rm me}(m,N) + Q_{\rm fp}(m,N)$ then denotes the probability that a walk
stays positive up to $(N-1)$ steps and has $m$ local minima. Now, a local minimum
may or may not form exactly at the $N$-th step. If the former happens, the meander walk
must have $(m-1)$ minima till the $(N-1)$-th step and a new minima occurs exactly
at the $N$-th step, contributing to the event of $m$ minima up to $N$ steps.
In the latter case, the meander walk must have already $m$ minima up to $(N-1)$ steps
and no new minimum forms after the $N$-th jump. In order to keep track of this
last-step event, we need to consider a meander walk up to $(N-1)$ steps with
the $(N-1)$-th jump either positive or negative (as shown respectively in the 
left and the right panel of Fig. (\ref{fp_updown.fig})). If the $(N-1)$-th step
is positive, then no matter what the sign of the $N$-th jump is, there is no
possibility of forming a new local minimum by the $N$-th jump (see the left panel
of Fig. (\ref{fp_updown.fig})). In contrast, if the $(N-1)$-th jump is negative,
then there are two possibilities: (i) if the $N$-th jump is positive (which
occurs with probability $1/2$), then a new minimum
forms exactly after the $N$-th jump  and (ii) if the $N$-th jump is negative (which also
occurs with probability $1/2$), no new minimum
is generated at the $N$-th step. Considering these possibilities, it is then easy to write down the
following exact recursion relation
\begin{equation}
Q_{\rm me}(m,N)+ Q_{\rm fp}(m,N)= Q_{\rm me}^{+}(m, N-1)+ \frac{1}{2}\, Q_{\rm me}^{-}(m-1,N-1)+\frac{1}{2}\, Q_{\rm me}^{-}(m,N-1)
\, ,
\label{recur_fp.1}
\end{equation}
where the first term on the rhs refers to the event when the $(N-1)$-th jump of the
meander is positive (in that case it has to have $m$ minima till $(N-1)$ steps since
no new minimum is generated at the $N$-th step), while the last two terms counts
the probability of events (i) and (ii) described above. Using Eq. (\ref{recur_fp.1})
and the definition $Q_{\rm me}(m,N)= Q_{\rm me}^{+}(m,N)+ Q_{\rm me}^{-}(m,N)$,
we can then express $Q_{\rm fp}(m,N)$ in terms of the known quantities $Q_{\rm me}^{\pm }(m,N)$
as
\begin{equation}
Q_{\rm fp}(m,N)= Q_{\rm me}^{+}(m,N-1) + \frac{1}{2}\, Q_{\rm me}^{-}(m,N-1)
+ \frac{1}{2}\, Q_{\rm me}^{-}(m-1,N-1) -  Q_{\rm me}^{+}(m,N)+ Q_{\rm me}^{-}(m,N)\, .
\label{recur_fp.2}
\end{equation}
This recursion relation actually holds for $m\ge 1$ and $N\ge 1$,
with the convention that $Q_{\rm me}(m,0)= \delta_{m,0}$ and $Q_{\rm me}^{-}(m,0)=0$.
For the special case $m=0$, one can
similarly write down the recursion relation
\begin{equation}
Q_{\rm fp}(0,N)= Q_{\rm me}^{+}(0, N-1) + \frac{1}{2}\, Q_{\rm me}^{-}(0,N-1)
+  \frac{1}{2}\, Q_{\rm me}^{-}(-1,N-1)
- Q_{\rm me}^+(0,N)- Q_{\rm me}^{-}(0,N)\, .
\label{m0_recur_fp.1}
\end{equation}
Thus, we can actually include the $m=0$ term in the general recursion relation (\ref{recur_fp.2})
valid for $m\ge 1$, provided we interpret $Q_{\rm me}^{-}(-1,N)=0$. From now on
we use this convention and hence Eq. (\ref{recur_fp.2}) is valid for all $m\ge 0$ and $N\ge 1$.

\begin{figure}
\centering
\includegraphics[width=0.6\linewidth]{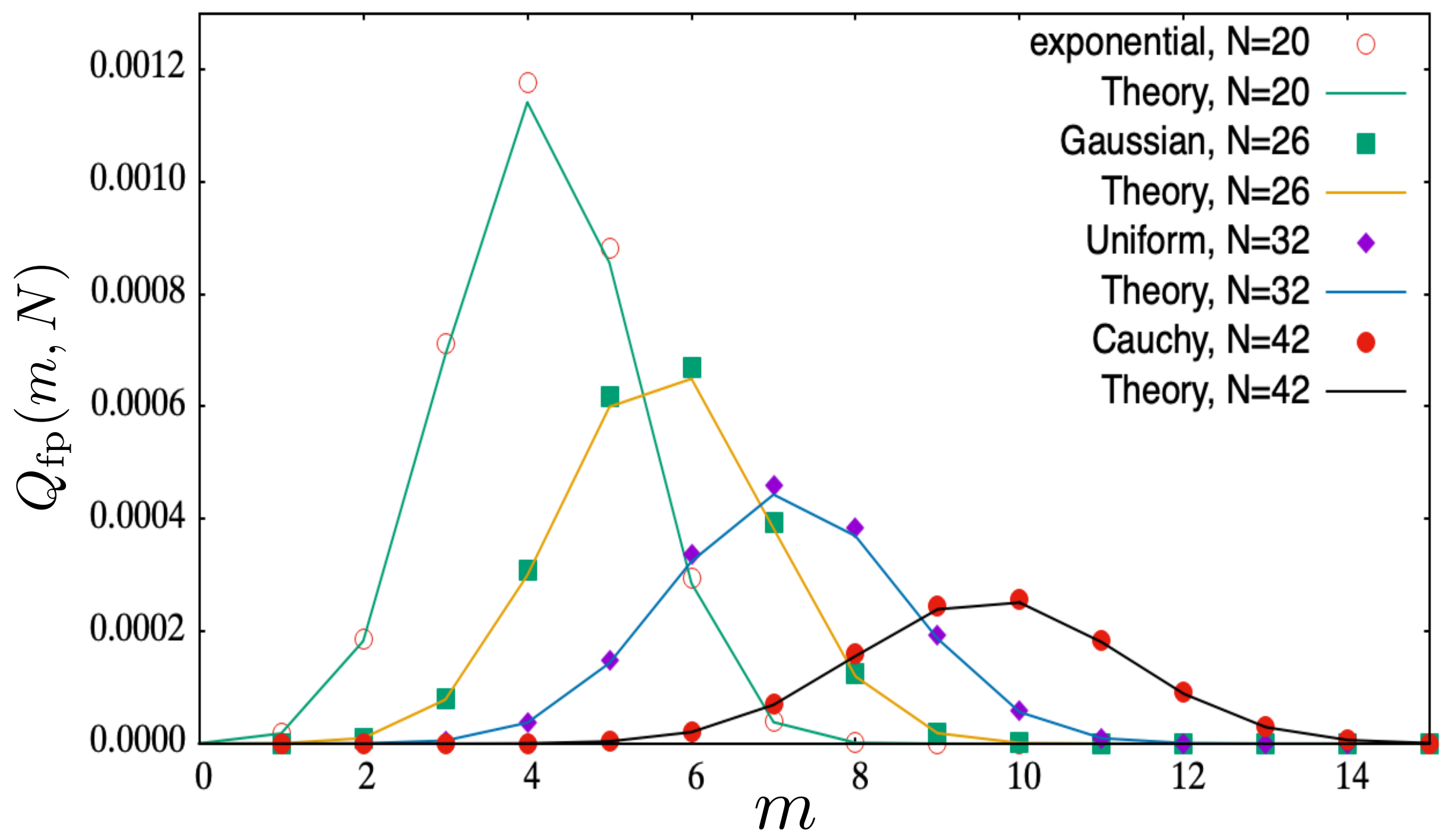}
\caption{Plot of the numerically obtained $Q_{\rm fp}(m,N)$ for four different jump
distributions and for four different values of $N$, compared to theoretical expression
in Eq.~\eqref{result.MIII} (shown by solid lines), showing a perfect agreement.}
\label{fp_num.fig}
\end{figure}

To proceed, we define the generating function
\begin{equation}
\tilde{Q}_{\rm fp}(m,z)= \sum_{N=1}^{\infty} Q_{\rm fp}(m,N)\, z^N\, .
\label{fp_genf_def}
\end{equation}
Taking generating function of Eq. (\ref{recur_fp.2}) and using the convention
$Q_{\rm me}^{+}(m,0)=\delta_{m,0}$ and $Q_{\rm me}^{-}(m,0)=0$, we get
\begin{equation}
\tilde{Q}_{\rm fp}(m,z)= z\, \delta_{m,0}+ (z-1)\, \tilde{Q}_{\rm me}^{+}(m,z)
+ \frac{z}{2}\, \tilde{Q}_{\rm me}^{-}(m,z)+ \frac{z}{2}\, \tilde{Q}_{\rm me}^{-}(m-1,z)
- \tilde{Q}_{\rm me}^{-}(m,z)\, ,
\label{recur_fp_genf.1}
\end{equation}
where $\tilde{Q}_{\rm me}(m,z)= \sum_{N=1}^{\infty} Q_{\rm me}^{\pm}(m, N)\, z^N$.
Using the explicit results for $\tilde{Q}_{\rm me}^{\pm}(z)$ derived
respectively in Eqs. (\ref{Pp_genf_me.3}) and (\ref{Pm_genf_me.3}), we get, after
a few simplifying steps,
\begin{eqnarray}
\label{sol_fp.1}
\tilde{Q}_{\rm fp}(m,z)= \begin{cases}
\frac{z}{2}\, \left(\frac{z}{2-z}\right)^{m+1}\,
\left(q_m- q_{m+1}\right)  \quad {\rm for} \quad m\ge 1 \\
\\
\frac{z(4-z)}{4\,(2-z)} \hspace{3.1cm} \quad {\rm for} \quad m=0\, ,
\end{cases} 
\end{eqnarray}
where $q_m$ is given in Eq. (\ref{SA_result.1}). Using this explicit result for $q_m$,
Eq. (\ref{sol_fp.1}) reduces to
\begin{eqnarray}
\label{sol_fp.2}
\tilde{Q}_{\rm fp}(m,z)= \begin{cases}
\frac{(2m)!}{2^{4m+3}\, m!\, (m+1)!}\, 
z^{2m+2}\, \left(1- \frac{z}{2}\right)^{-(2m+1)}\, \quad {\rm for}\quad m\ge 1 \\
\\
\frac{z(4-z)}{4\,(2-z)} \hspace{4.8cm} \quad {\rm for} \quad m=0\, .
\end{cases}
\end{eqnarray}
Let us remark that by setting $z=1$ in (\ref{sol_fp.2}), and using 
$Q^{\rm fp}(m)= \sum_{N=1}^{\infty} Q_{\rm fp}(m, N)$, we obtain
\begin{eqnarray}
\label{fp_total.1}
Q^{\rm fp}(m)= \begin{cases}
\frac{1}{2^{2m+2}}\, \frac{(2m)!}{m!\, (m+1)!}\, \quad {\rm for} \quad m\ge 1 \\
\\
\frac{3}{4} \quad\quad \hspace{1.8cm}{\rm for} \quad m=0\, ,
\end{cases}
\end{eqnarray}
thus recovering the result that was derived in Ref.~\cite{KMS24} by a slightly different method.
We next use the identity (\ref{iden.2}) to expand the rhs in powers of $z$ and then match
the powers of $z$ on both sides of Eq. (\ref{sol_fp.2}). This then gives the desired result
(\ref{mdist.III}), namely, for all $N\ge 2$
\begin{eqnarray}
\label{result.MIII}
Q_{\rm fp}(m,N)  = \begin{cases} \frac{ (N-1)!}{2^{2m+N+1}\, m!\, (m+1)!\, \Gamma(N-2m-1)}
\quad {\rm for}\quad 0\le m\le \frac{N}{2}-1\, ,
   \\
\\
0 \hspace{4.4cm} {\rm for}\quad m > \frac{N}{2}-1\, ,
\end{cases}
\end{eqnarray}
and for $N=1$, we get $Q_{\rm fp}(m,1)= \frac{1}{2}\, \delta_{m,0}$.
These analytical predictions are compared to numerical simulations in Fig. (\ref{fp_num.fig}),
finding excellent agreement.

\section{Summary and Conclusion} \label{conclusion}

In this paper we have provided a unified framework to compute the exact distribution of the number of minima/maxima in three one-dimensional random walk landscapes. The first landscape corresponds to the trajectory of a free discrete-time random walk of $N$ steps with arbitrary symmetric jump distribution at each step. The second ``meander landscape'' model corresponds to the trajectory a constrained random walk that starts at the origin and remains non-negative up to step $N$. In the third ``first-passage landscape '' we consider a random walk trajectory that starts at the origin and stops when it crosses the origin for the first time. Unlike in the first two models, here the number of steps of the landscape is a random variable that fluctuates from sample to sample. Our main result is to show that while the exact distribution of the number of minima is different in the three models, for each model it is universal for all $N$, in the sense that it does not depend on the jump distribution as long as it is symmetric and continuous. In the last two cases we show that this universality follows from a deep connection to the Sparre Andersen theorem known for the first-passage probability of discrete-time random walks with symmetric and continuous jump distribution. Our analytical results are in excellent agreement with our numerical simulations.

There are many directions in which this work can be extended. For example, in Ref. \cite{KMS24}, we showed that, for a free random walk landscape the
joint distribution of the number of minima and maxima up to $N$ steps is also universal and can be computed explicitly. It would be interesting to compute this joint distribution for the other two constrained random walk landscape models discussed in this paper and see whether there are also universal. {The unified framework developed here via the mapping to the auxiliary random walk may be possible to use to compute the distribution of the number of minima or maxima (sometimes called ``peaks'') for other constrained random walks, such as the random walk bridge or random walk excursion, as it was done for lattice random walks \cite{Marckert}.} Finally, it would be interesting to extend these studies of the statistics of the number of local minima, maxima and saddles to correlated landscapes in higher dimensions.

\acknowledgments
The authors would like to thank the Isaac Newton Institute for Mathematical Sciences, Cambridge, for support and hospitality during the programmes 
{\it New statistical physics in living matter: non equilibrium states under adaptive control} and {\it Stochastic systems for anomalous diffusion}. 
where work on this paper was undertaken. This work was supported by by EPSRC Grant Number EP/R014604/1 and EPSRC grant no EP/K032208/1. SNM and GS acknowledge support from ANR Grant No. ANR-23-CE30-0020-01 EDIPS. SNM acknowledges the support from the Science and Engineering Research Board (SERB, Government of India), under the VAJRA faculty scheme (No. VJR/2017/000110). AK would like to acknowledge the support of DST, Government of India Grant under Project No. CRG/2021/002455 and the MATRICS Grant No. MTR/2021/000350 from the SERB, DST, Government of India. AK acknowledges the Department of Atomic Energy, Government of India, for their support under Project No. RTI4001.



%


\begin{thebibliography}{15}%
\makeatletter
\providecommand \@ifxundefined [1]{%
 \@ifx{#1\undefined}
}%
\providecommand \@ifnum [1]{%
 \ifnum #1\expandafter \@firstoftwo
 \else \expandafter \@secondoftwo
 \fi
}%
\providecommand \@ifx [1]{%
 \ifx #1\expandafter \@firstoftwo
 \else \expandafter \@secondoftwo
 \fi
}%
\providecommand \natexlab [1]{#1}%
\providecommand \enquote  [1]{``#1''}%
\providecommand \bibnamefont  [1]{#1}%
\providecommand \bibfnamefont [1]{#1}%
\providecommand \citenamefont [1]{#1}%
\providecommand \href@noop [0]{\@secondoftwo}%
\providecommand \href [0]{\begingroup \@sanitize@url \@href}%
\providecommand \@href[1]{\@@startlink{#1}\@@href}%
\providecommand \@@href[1]{\endgroup#1\@@endlink}%
\providecommand \@sanitize@url [0]{\catcode `\\12\catcode `\$12\catcode
  `\&12\catcode `\#12\catcode `\^12\catcode `\_12\catcode `\%12\relax}%
\providecommand \@@startlink[1]{}%
\providecommand \@@endlink[0]{}%
\providecommand \url  [0]{\begingroup\@sanitize@url \@url }%
\providecommand \@url [1]{\endgroup\@href {#1}{\urlprefix }}%
\providecommand \urlprefix  [0]{URL }%
\providecommand \Eprint [0]{\href }%
\@ifxundefined \urlstyle {%
  \providecommand \doi  [0]{\begingroup \@sanitize@url \@doi}%
  \providecommand \@doi [1]{\endgroup \@@startlink {\doibase
  #1}doi:\discretionary {}{}{}#1\@@endlink }%
}{%
  \providecommand \doi  [0]{doi:\discretionary{}{}{}\begingroup
  \urlstyle{rm}\Url }%
}%
\providecommand \doibase [0]{http://dx.doi.org/}%
\providecommand \Doi [0]{\begingroup \@sanitize@url \@Doi }%
\providecommand \@Doi  [1]{\endgroup\@@startlink{\doibase#1}\@@Doi}%
\providecommand \@@Doi [1]{#1\@@endlink}%
\providecommand \selectlanguage [0]{\@gobble}%
\providecommand \bibinfo  [0]{\@secondoftwo}%
\providecommand \bibfield  [0]{\@secondoftwo}%
\providecommand \translation [1]{[#1]}%
\providecommand \BibitemOpen [0]{}%
\providecommand \bibitemStop [0]{}%
\providecommand \bibitemNoStop [0]{.\EOS\space}%
\providecommand \EOS [0]{\spacefactor3000\relax}%
\providecommand \BibitemShut  [1]{\csname bibitem#1\endcsname}%

\bibitem{Adler}
R. J. Adler, J. E. Taylor, {\it Random Fields and Geometry} (Berlin: Springer), (2009)

\bibitem{Azais}
J.-M. Aza\"is, M. Wschebor, {\it Level Sets and Extrema of Random Processes and Fields}, (New
York: Wiley), (2009)



\bibitem [{\citenamefont {Freund}(1995)}]{freund1995saddles}%
  \BibitemOpen
  \bibfield  {author} {\bibinfo {author} {\bibfnamefont {I.}~\bibnamefont
  {Freund}},\ }\href@noop {} {\bibfield  {journal} {\bibinfo  {journal}
  {Phys. Rev. E},\ }\textbf {\bibinfo {volume} {52}},\ \bibinfo {pages}
  {2348} (\bibinfo {year} {1995})}\BibitemShut {NoStop}%
\bibitem [{\citenamefont {Halperin}\ and\ \citenamefont
  {Lax}(1966)}]{halperin1966impurity}%
  \BibitemOpen
  \bibfield  {author} {\bibinfo {author} {\bibfnamefont {B.}~\bibnamefont
  {Halperin}}\ and\ \bibinfo {author} {\bibfnamefont {M.}~\bibnamefont {Lax}},\
  }\href@noop {} {\bibfield  {journal} {\bibinfo  {journal} {Phys. Rev.},\
  }\textbf {\bibinfo {volume} {148}},\ \bibinfo {pages} {722} (\bibinfo {year}
  {1966})}\BibitemShut {NoStop}%
\bibitem [{\citenamefont {Broderix}\ \emph {et~al.}(2000)\citenamefont
  {Broderix}, \citenamefont {Bhattacharya}, \citenamefont {Cavagna},
  \citenamefont {Zippelius},\ and\ \citenamefont
  {Giardina}}]{broderix2000energy}%
  \BibitemOpen
  \bibfield  {author} {\bibinfo {author} {\bibfnamefont {K.}~\bibnamefont
  {Broderix}}, \bibinfo {author} {\bibfnamefont {K.~K.}\ \bibnamefont
  {Bhattacharya}}, \bibinfo {author} {\bibfnamefont {A.}~\bibnamefont
  {Cavagna}}, \bibinfo {author} {\bibfnamefont {A.}~\bibnamefont {Zippelius}},
  \ and\ \bibinfo {author} {\bibfnamefont {I.}~\bibnamefont {Giardina}},\
  }\href@noop {} {\bibfield  {journal} {\bibinfo  {journal} {Phys. Rev.
  Lett.},\ }\textbf {\bibinfo {volume} {85}},\ \bibinfo {pages} {5360}
  (\bibinfo {year} {2000})}\BibitemShut {NoStop}%
\bibitem [{\citenamefont {Longuet-Higgins}(1960)}]{longuet1960reflection}%
  \BibitemOpen
  \bibfield  {author} {\bibinfo {author} {\bibfnamefont {M.}~\bibnamefont
  {Longuet-Higgins}},\ }\href@noop {} {\bibfield  {journal} {\bibinfo
  {journal} {JOSA},\ }\textbf {\bibinfo {volume} {50}},\ \bibinfo {pages} {845}
  (\bibinfo {year} {1960})}\BibitemShut {NoStop}%
  %
 \bibitem{halperin_82}
 A. Weinrib, B. I. Halperin, Phys. Rev. B {\bf 26}, 1362 (1982).
 %
 \bibitem{Ros_Fyo_23}
V. Ros, Y. V.  Fyodorov, in {\it The High-dimensional Landscape Paradigm: Spin-Glasses, and Beyond. In Spin Glass Theory and Far Beyond: Replica Symmetry Breaking After 40 Years}, arXiv:2209.07975, (2023). 


\bibitem [{\citenamefont {Bray}\ and\ \citenamefont
  {Moore}(1980)}]{bray1980metastable}%
  \BibitemOpen
  \bibfield  {author} {\bibinfo {author} {\bibfnamefont {A.~J.}\ \bibnamefont
  {Bray}}\ and\ \bibinfo {author} {\bibfnamefont {M.~A.}\ \bibnamefont
  {Moore}},\ }\href@noop {} {\bibfield  {journal} {\bibinfo  {journal} {J. Phys. C: Solid State Phys.},\ }\textbf {\bibinfo {volume} {13}},\
  \bibinfo {pages} {L469} (\bibinfo {year} {1980})}\BibitemShut {NoStop}%
\bibitem [{\citenamefont {Annibale}\ \emph {et~al.}(2003)\citenamefont
  {Annibale}, \citenamefont {Cavagna}, \citenamefont {Giardina},\ and\
  \citenamefont {Parisi}}]{annibale2003supersymmetric}%
  \BibitemOpen
  \bibfield  {author} {\bibinfo {author} {\bibfnamefont {A.}~\bibnamefont
  {Annibale}}, \bibinfo {author} {\bibfnamefont {A.}~\bibnamefont {Cavagna}},
  \bibinfo {author} {\bibfnamefont {I.}~\bibnamefont {Giardina}}, \ and\
  \bibinfo {author} {\bibfnamefont {G.}~\bibnamefont {Parisi}},\ }\href@noop {}
  {\bibfield  {journal} {\bibinfo  {journal} {Phys. Rev. E},\ }\textbf
  {\bibinfo {volume} {68}},\ \bibinfo {pages} {061103} (\bibinfo {year}
  {2003})}\BibitemShut {NoStop}%
\bibitem [{\citenamefont {Aspelmeier}\ \emph {et~al.}(2004)\citenamefont
  {Aspelmeier}, \citenamefont {Bray},\ and\ \citenamefont
  {Moore}}]{aspelmeier2004complexity}%
  \BibitemOpen
  \bibfield  {author} {\bibinfo {author} {\bibfnamefont {T.}~\bibnamefont
  {Aspelmeier}}, \bibinfo {author} {\bibfnamefont {A. J.}~\bibnamefont {Bray}}, \
  and\ \bibinfo {author} {\bibfnamefont {M.}~\bibnamefont {Moore}},\
  }\href@noop {} {\bibfield  {journal} {\bibinfo  {journal} {Phys. Rev.
  Lett.},\ }\textbf {\bibinfo {volume} {92}},\ \bibinfo {pages} {087203}
  (\bibinfo {year} {2004})}\BibitemShut {NoStop}%
 %
 \bibitem{Satya_Martin}
S. N. Majumdar, O. C. Martin, Phys. Rev. E {\bf 74}, 061112 (2006). 
%
  %
 \bibitem{Hivert}
F. Hivert, S. Nechaev, G. Oshanin, O. Vasilyev, J. Stat. Phys. {\bf 126}, 243 (2007). 
\bibitem{Sollich}  
P. Sollich, S. N. Majumdar, A. J. Bray, J. Stat. Mech., 11011 (2008).
%
\bibitem [{\citenamefont {Aazami}\ and\ \citenamefont
  {Easther}(2006)}]{aazami2006cosmology}%
  \BibitemOpen
  \bibfield  {author} {\bibinfo {author} {\bibfnamefont {A.}~\bibnamefont
  {Aazami}}\ and\ \bibinfo {author} {\bibfnamefont {R.}~\bibnamefont
  {Easther}},\ }\href@noop {} {\bibfield  {journal} {\bibinfo  {journal}
  {J. Cosmo. Astr. Phys.},\ }\textbf {\bibinfo
  {volume} {2006}},\ \bibinfo {pages} {013} (\bibinfo {year}
  {2006})}\BibitemShut {NoStop}%
\bibitem [{\citenamefont {Susskind}(2003)}]{susskind2003anthropic}%
  \BibitemOpen
  \bibfield  {author} {\bibinfo {author} {\bibfnamefont {L.}~\bibnamefont
  {Susskind}},\ }\href@noop {} {\bibfield  {journal} {\bibinfo  {journal}
  {arXiv preprint hep-th/0302219}} (\bibinfo {year} {2003})}\BibitemShut
  {NoStop}%
%
\bibitem [{\citenamefont {Barton}(2005)}]{barton2005fitness}%
  \BibitemOpen
  \bibfield  {author} {\bibinfo {author} {\bibfnamefont {N.}~\bibnamefont
  {Barton}},\ }\href@noop {} {\enquote {\bibinfo {title} {Fitness landscapes
  and the origin of species},}\ } (\bibinfo {year} {2005})\BibitemShut
  {NoStop}%
 %
 \bibitem{Krug_review}
I. G. Szendro, M. F. Schenk, J. Franke, J. Krug, J. A. G. De Visser, J. Stat. Mech. 01005 (2013).
  
 \bibitem{Krug_fitness1} 
  S.-C. Park, S. Hwang, J. Krug, J. Phys. A: Math. Theor. {\bf 53}, 385601 (2020).  
  
 \bibitem{Krug_fitness2} 
K. Crona, J. Krug, M. Srivastava, J. Math. Bio. {\bf 86}, 62 (2023).

 \bibitem{Ganguli_14} 
Y. N. Dauphin, R. Pascanu, C. Gulcehre, K. Cho, S. Ganguli, Y. Bengio, Adv. Neur. In. {\bf 27} (2014).  

\bibitem{Rice}
 S. O. Rice, in {\it Selected Papers on Noise and Stochastic Processes}, edited by N. Wax (Dover, New York, 1954). 
  

\bibitem{KMS24} A. Kundu, S, N. Majumdar, G. Schehr, {\em Universal distribution of the number of minima for random walks and L\'evy flights}, Phys. Rev. E, {\bf 110}, 024137 (2024). 


\bibitem{SA1954} E. Sparre Andersen, Math. Scand. {\bf 2}, 195 (1954). 



\bibitem{Marckert}
J. - M. Labarbe, J. - F. Marckert, Electron. J. Probab. {\bf 12}, 229 (2007).   



  
\end{thebibliography}
\end{document}